\documentclass[a4paper,10pt,twoside,onecolumn]{article}
\usepackage[left=1.5cm, right=1.5cm, top=1.5cm, bottom=1.5cm]{geometry}

\usepackage[numbers,super]{natbib}
\usepackage[utf8]{inputenc}
\usepackage[T1]{fontenc}

\usepackage{fourier}

\usepackage[english]{babel}
\usepackage[hidelinks]{hyperref}

\usepackage{amsmath}
\allowdisplaybreaks
\usepackage{amsfonts}
\usepackage{bm}
\usepackage{mathrsfs}
\usepackage{booktabs}
\usepackage{graphicx}
\usepackage{url}
\usepackage{doi}
\usepackage{xcolor}

\usepackage{amsthm}
\theoremstyle{remark}
\newtheorem*{remark}{Remark}

\begin{document}

	\title{Computation of viscoelastic shear shock waves \\ using finite volume schemes with artificial compressibility}
	
	\author{Harold Berjamin \\[2pt]
	\emph{School of Mathematical and Statistical Sciences}, \\
	\emph{University of Galway, University Road, Galway, Ireland.} \\ \href{mailto:harold.berjamin@universityofgalway.ie}{harold.berjamin@universityofgalway.ie}}
	
	\date{}
	
	\maketitle
	
	\begin{abstract}
		\noindent
		The formation of shear shock waves in the brain has been proposed as one of the plausible explanations for deep intracranial injuries. In fact, such singular solutions emerge naturally in soft viscoelastic tissues under dynamic loading conditions. To improve our understanding of the mechanical processes at hand, the development of dedicated computational models is needed. The present study concerns three-dimensional numerical models of incompressible viscoelastic solids whose motion is analysed by means of shock-capturing finite volume methods. More specifically, we focus on the use of the artificial compressibility method, a technique that has been frequently employed in computational fluid dynamics. The material behaviour is deduced from the Fung--Simo quasi-linear viscoelasiticity theory (QLV) where the elastic response is of Yeoh type. We analyse the accuracy of the method and demonstrate its applicability for the study of nonlinear wave propagation in soft solids. The numerical results cover accuracy tests, shock formation and wave focusing.\\
		
	\end{abstract}

\section{Introduction}\label{sec:Intro}

Understanding the causes of traumatic brain injury (including mild traumatic brain injury, aka concussion) is very challenging. Due to obvious experimental constraints, theoretical and computational modelling of the motion of brain tissue in traumatic situations is a promising venue to progress towards this goal. In this context, a model based on laboratory experiments was used to demonstrate shear shock formation in soft viscoelastic solids such as brain tissue and gelatin \cite{giammarinaro16,tripathi19,tripathi19b}, thus providing a plausible explanation of deep traumatic brain injury.

The development of suitable computer models relies on accurate experiment-based mechanical models, which would take into account the main mechanical properties of soft solids, summarised hereinafter. Due to a large contrast between compression and shear stiffness, brain tissue can be considered incompressible. Moreover, brain tissue exhibits time-dependent viscoelastic behaviour, see the review by Budday et al. \cite{budday20} for complements. Therefore, up to moderate but finite strains and frequencies, quasi-linear viscoelastic (QLV) solid models are able to reproduce the main features of the mechanical response in an adequate fashion, see for instance Rashid et al. \cite{rashid13}.

Such an incompressible Fung or Simo model was used by Berjamin and Chockalingam \cite{berjamin21b} to solve one-dimensional shear wave propagation problems with formation of shocks. Under certain conditions, these discontinuous wave profiles develop naturally in finite time due to the nonlinearity of the constitutive law. If no special care is taken, the formation of shock waves can lead to the emergence of oscillations near strong gradients, a common numerical artifact caused by the singularity of the solution. In the above-mentioned study, this issue was tackled by using a Lagrangian Godunov-type finite volume method with limiters.

Aiming for the resolution of incompressible three-dimensional problems, let us first briefly introduce several computational approaches from the elasticity literature. In solid dynamics, the Lagrangian specification of motion is often used, i.e., the spatial coordinate is the position $\bm{X}$ of a particle in the undeformed configuration{\,---\,}as opposed to the Eulerian position $\bm{x}$ of a particle in the deformed configuration. The equations of motion can be recast as a first-order system of balance laws, where the primary variables are the deformation gradient tensor ${\bm F} = \partial\bm{x}/\partial\bm{X}$ and the particle velocity $\dot{\bm x}$, defined as the material time derivative of the current position (overdot). In this framework, Godunov-type finite volume methods appear as a viable option to efficiently solve nonlinear solid dynamics problems, more specifically when shock waves form \cite{cardiff21}.

Perfect \emph{incompressibility} requires that the condition $J \equiv 1$ is satisfied where $J = \det \bm{F}$ is the volume dilation. It follows that a Lagrangian multiplier $p$ for this constraint is introduced, whose physical dimension is that of pressure. Thus, compared to the unconstrained compressible case, the numerical resolution of such a problem requires the computation of an additional scalar quantity. This issue arises in general three-dimensional settings, but not in the special case of one-dimensional and two-dimensional simple shearing motions \cite{berjamin21b,tripathi19}.

Incompressibility is addressed in the literature based on various strategies. For a specific class of finite volume methods \cite{cardiff21}, Bijelonja et al. \cite{bijelonja05} discretise the integral balance of momentum in its second-order form, as well as the integral version of the incompressibility constraint $\dot J \equiv 0$ obtained through time-differentiation of the initial restriction. Several computational strategies have also been proposed in the Finite Element literature, including mixed methods and stabilised methods \cite{gil16, rossi16, castanar20, castanar23}, as well as penalisation methods \cite{caforio18}, including computational models that account for viscoelastic material behaviour \cite{simo87,liu21}. However, the issue of incompressibility remains to be further explored in the context of Lagrangian Godunov-type finite volume methods, especially in relation with the viscoelastic behaviour of soft tissues, and their nonlinear response to transient loadings.

Following Lee et al. \cite{lee13}, we derive and implement an \emph{artificial compressibility} (AC) method based on the approximate incompressibility constraint $\dot{J} = -\dot{p}/K$ where $K$ is a large stiffness parameter and $p$ is the hydrostatic pressure (see also Lee's thesis \cite{lee12}). Here, this process amounts to the replacement of the truly incompressible model through a nearly incompressible one. In effect, the artificial compressibility method is a perturbation of the perfectly incompressible theory that takes into account a moderate level of compressibility, and whose associated errors are transported out of the computational domain with a high artificial velocity.

Historically, the artificial compressibility method originated in computational fluid dynamics in the mid-1960s, where it was proposed to solve steady incompressible flow problems \cite{chorin97,drikakis05}. In the dedicated literature, the AC method was extended to unsteady incompressible flow by introducing a pseudo-time such that perfect incompressibility can be viewed as a pseudo-equilibrium. Then, dual time-stepping is performed, that is, integration in physical time and in pseudo-time are alternated. The AC equations can also be integrated directly in physical time \cite{madsen06}, up to a careful choice of the AC parameter $K$. Classical alternatives to artificial compressibility include so-called `pressure-Poisson' and `projection'-based predictor-corrector schemes \cite{brown01}.

As a side note, the AC method bears some similarity with the artificial viscosity approach \cite{tripathi18,margolin19}. In both cases, suitable choices of the small parameter result from a compromise between advantageous numerical properties and good physical accuracy. Moreover, despite being called `artificial', these asymptotic methods are actually physics-related. A major difference is the nature of the physical process introduced by means of an extra parameter. In fact, the AC method introduces unconstrained compressible behaviour (thus circumventing strict incompressibility), whereas the artificial viscosity regularisation introduces material dissipation leading to smoother solutions.

Beyond the enforcement of incompressibility, we aim for the computation of shock waves, which are singular time-dependent solutions that can arise in nonlinear viscoelasticity, typically in dynamic problems involving rubber-like solids and soft tissues. In this context, the use of classical high-order finite difference methods requires special care for the treatment of shocks, either via a careful choice of the mesh size that allows to resolve very small wavelengths \cite{cuenca20}, or by making use of dedicated shock-capturing filters \cite{bogey09,sabatini16}. Furthermore, classical finite elements methods based on a weak formulation of the equations of motion have shown their limitations as well. In fact, more sophisticated techniques have been proposed to accommodate for fast transient solutions, see for instance the dedicated procedure proposed by Wellford and Oden \cite{wellford75} which involves finite elements with built-in discontinuities, as well as the literature on discontinuous Galerkin methods \cite{bertoluzza09, boumatar12}. In solid dynamics, a few recent studies have produced satisfactory results for the simulation of shocks using finite element techniques, see for instance Renaud et al. \cite{renaud18} where discontinuous Galerkin approximations are exploited.

Originating in fluid dynamics, Godunov-type finite volume methods are both easy to implement and well-suited to the numerical resolution of \emph{balance laws}, including the computation of shock waves \cite{godlewski96,leveque02,toro09}. Consequently, the implementation of these algorithms represents a relevant step forward in the present context of Lagrangian solid dynamics \cite{lee13}. Formally, finite volume methods have similar features to finite element methods, such as the possibility of using them on structured or unstructured meshes, and their robustness. Moreover, they can be reinterpreted as a special class of discontinuous finite element methods, for a peculiar choice of shape and test functions \cite{morton07}. Finite volume methods were used by several authors in the solid dynamics context to compute nonlinear waves, based on the piecewise parabolic method \cite{tripathi19,tripathi19b}, MUSCL schemes with limiters \cite{berjamin21b}, or a Godunov scheme \cite{favrie23}. Successful extensions to high order on unstructured meshes with adaptive refinement have been proposed \cite{dumbser13,exahype}, and their potential use for the computation of fast motions in solid dynamics is promising.

In the present paper, we give a detailed overview of the artificial compressibility method in Lagrangian solid dynamics, where the method is applied to a family of Godunov-type finite volume schemes. This computational framework is used for studying the motion of soft viscoelastic tissues in multiple spatial dimensions, where the effects of material nonlinearity, dissipation, and geometry are combined \cite{tripathi19}. Special attention is given to the computation of shear shock waves, which is enabled based on the use of the MUSCL--Hancock slope limiting procedure \cite{toro09}.

The paper is organised as follows. Section~\ref{sec:GovEq} introduces the equations governing the motion of incompressible quasi-linear viscoelastic solids (QLV). The model is adapted to match an existing model of gelatin \cite{tripathi19} by using a two-term Yeoh strain energy function for the material's elastic response, while the time-dependent material response follows Simo's viscoelasticity theory \cite{simo87}. Numerical resolution of the equations of motion is addressed in Section~\ref{sec:FV}. Here, we present a simple Godunov-type finite volume method based on the artificial compressibility technique, MUSCL reconstruction and an approximate Riemann solver. Numerical results are shown in Section~\ref{sec:Results}, where we demonstrate the applicability of the method. This way, the one-dimensional \cite{tripathi19b,berjamin21b} and two-dimensional results \cite{tripathi19} found in the literature are extended to three spatial dimensions.

\section{Governing equations}\label{sec:GovEq}

\subsection{Finite deformations}\label{ssec:Def}

In what follows, we present the basic equations of Lagrangian dynamics for incompressible solids \cite{holzapfel00}. We consider a homogeneous and isotropic solid continuum on which no external body force is applied. The deformation gradient tensor is the second-order tensor
\begin{equation}
	\bm F = \frac{\partial{\bm{x}}}{\partial{\bm{X}}} = \bm{I} + \nabla \bm{u} ,
	\label{F}
\end{equation}
where $\bm{u} = \bm{x} - \bm{X}$ is the displacement field and $\nabla \bm{u} = \partial\bm{u}/\partial\bm{X}$ is the displacement gradient tensor in material coordinates. If the Euclidean space is described by an orthonormal basis $\lbrace \bm{e}_1, \bm{e}_2, \bm{e}_3 \rbrace$ and a Cartesian coordinate system, then the metric tensor $\bm{I}$ has Kronecker delta components $[\delta_{ij}]$. Various finite strain tensors are defined as functions of $\bm F$, such as the left and right Cauchy--Green tensors $\bm{B} = \bm{F} \bm{F}^{\top}$, $\bm{C} = \bm{F}^{\top} \bm{F}$, as well as the Green--Lagrange tensor $\bm{E} = \frac{1}{2} (\bm{C} - \bm{I})$.

Here, we consider \emph{incompressible} materials, for which the deformation gradient tensor $\bm F$ is unimodular. Indeed, the constraint of no volume dilatation
\begin{equation}
	J = \det\bm{F} \equiv 1
	\label{Dilat}
\end{equation}
is prescribed at all times. We note in passing that the deformation tensors $\bm{B}$, $\bm C$ defined as the products between the deformation gradient and its transpose have unit determinant too. The constraint \eqref{Dilat} implies also that the mass density $\rho$ is constant. In other words, it equals the initial value $\rho = \rho_0$ in the undeformed configuration.

Through time differentiation of \eqref{F}, we have
\begin{equation}
	\dot{\bm F} = \nabla \bm{v} ,
	\label{ConsF}
\end{equation}
where $\bm{v} = \dot{\bm u} = \dot{\bm x}$ is the Lagrangian velocity field, and the overdot denotes the material time derivative $\partial_t$. Furthermore, the incompressibility condition may be rewritten in differential form as \cite{holzapfel00}
\begin{equation}
	\dot{J} = J\bm{F}^{-\top} : \dot{\bm F} = \nabla\cdot \big(J\bm{F}^{-1}\bm{v}\big) \equiv 0 .
	\label{IncompDiff}
\end{equation}
Here, we have used several tensor calculus identities, including the Piola identity $\nabla\cdot (J\bm{F}^{-\top}) = \bm{0}$ and classical differentiation rules.
The motion is also governed by the conservation of momentum equation
\begin{equation}
	\rho_0 \dot{\bm v} = \nabla\cdot \bm{P} .
	\label{ConsMom}
\end{equation}
This equation of motion involves the divergence of the first Piola--Kirchhoff stress tensor $\bm P$ whose expression is specified later on. In what follows, we use the definition $[\nabla\cdot \bm{P}]_i = P_{ij,j}$ of the divergence operator, where summation over repeated indices is assumed \cite{holzapfel00}.

\subsection{Incompressible material model}\label{ssec:Visco}

In this section, we present the incompressible version of the theory used by De Pascalis et al. \cite{depascalis14} to model time-dependent viscoelastic deformations of soft isotropic solids. This model belongs to the quasi-linear viscoelastic (QLV) material models which combine finite elastic deformations at short and long times with linear relaxation mechanisms. The model amounts to a nonlinear viscoelasticity theory with memory variables. A similar theory accounting for material anisotropy can be found in the literature \cite{balbi23}.

In \emph{incompressible} and isotropic elastic solids, the strain energy function $W$ depends on two scalar invariants only, for instance the principal invariants $I_1 = \text{tr}\, \bm{C}$ and $I_2 = \frac12 \big(I_1^2 - \text{tr} (\bm{C}^2)\big)$. The resulting second Piola--Kirchhoff stress $\bm{S}^\text{e} = 2\, \partial W/\partial\bm{C}$ can be expressed as
\begin{equation}
	\bm{S}^\text{e} = 2 \left(W_1 + I_1 W_2\right) \bm{I} - 2W_2 \bm{C} ,
	\qquad\text{or}\qquad
	{\bm S}^\text{e}_\text{D} = \text{Dev}(\bm{S}^\text{e}) = \bm{S}^\text{e} - \tfrac23 \left(I_1 W_1 + 2 I_2 W_2\right) \bm{C}^{-1} ,
	\label{Elast}
\end{equation}
where $W_i$ is shorthand for the partial derivative $\partial W/\partial I_i$ evaluated at $(I_1, I_2)$. The deviatoric elastic stress ${\bm S}^\text{e}_\text{D}$ is obtained by application of the deviatoric operator $\text{Dev}(\bullet) = (\bullet) - \tfrac13 (\bullet :\bm{C}) \bm{C}^{-1}$ in the Lagrangian description \cite{holzapfel00}. Both stress tensors $\bm{S}^\text{e}$ and ${\bm S}^\text{e}_\text{D}$ are symmetric.

In incompressible QLV solids, the total second Piola--Kirchhoff stress $\bm{S} = \bm{F}^{-1}\bm{P}$ can be expressed as \cite{berjamin21b}
\begin{equation}
	\bm{S} = -q \bm{C}^{-1} + {\bm S}^\text{e} - \sum_{\ell=1}^N \bm{S}^\text{v}_\ell , \qquad
	\bm{S}^\text{v}_\ell = g_\ell\omega_\ell \int_0^t \text{e}^{-\omega_\ell (t-s)}\, {\bm S}^\text{e}_\text{D}(s)\, \text d s ,
	\label{Constitutive}
\end{equation}
where $q(\bm{X},t)$ is an arbitrary Lagrange multipliers accounting for the incompressibility constraint. The $N$ tensors $\bm{S}^\text{v}_\ell$ are stress-like \emph{memory variables} governed by linear evolution equations
\begin{equation}
	\dot{\bm S}^\text{v}_\ell = \omega_\ell \left(g_\ell {\bm S}^\text{e}_\text{D} - \bm{S}^\text{v}_\ell \right) , \quad 1 \leqslant \ell\leqslant N,
	\label{MemEvol}
\end{equation}
with parameters $g_\ell$, $\omega_\ell$. Here, the coefficients $\omega_\ell$ (in s\textsuperscript{$-1$}) are relaxation frequencies, whereas the dimensionless coefficients $g_\ell$ govern the magnitude of the $\ell$-th relaxation mechanism. Other equivalent expressions of the constitutive law can be obtained with suitable redefinitions of the Lagrange multiplier \cite{berjamin22c}.

Two elastic limits are identified. On the one hand, in the high frequency range where $\omega_\ell \to 0$, we note that ${\bm S}^\text{v}_\ell = \bm{0}$ for all $\ell$. In other words, the motion is too fast for the relaxation mechanisms, so that the memory variables remain equal to their initial value. On the other hand, in the low frequency range where $\omega_\ell \to +\infty$, then ${\bm S}^\text{v}_\ell = g_\ell {\bm S}^\text{e}_\text{D}$ for all $\ell$. Here, the memory variables are in a state of quasi-equilibrium. In both cases, the resulting stress \eqref{Constitutive} is purely elastic, but with distinct elastic responses, in general.

In practice, the fourth-order strain energy function
\begin{equation}
	W^\text{FOE} = \mu \operatorname{tr}(\bm{E}^{2}) + \tfrac13 \mathscr{A} \operatorname{tr}(\bm{E}^{3}) + \mathscr{D} \operatorname{tr}(\bm{E}^{2})^2 ,
	\label{WZabo}
\end{equation}
introduced by Zabolotskaya et al. \cite{zabo04} might be used, where $\mu > 0$ is the shear modulus, and the coefficients $\mathscr{A}$, $\mathscr{D}$ are the third- and fourth-order elastic constants. This approach was followed by Tripathi et al. \cite{tripathi19} based on a slightly different rheological model which involves an additive decomposition of the deformation gradient tensor instead of the second Piola--Kirchhoff stress \eqref{Constitutive}. The above strain energy function \eqref{WZabo} is equivalent to the combined Mooney--Rivlin and Yeoh model
\begin{equation}
	W^\text{MRY} = \mathscr{C}_{10} \left(I_1-3\right) + \mathscr{C}_{01} \left(I_2-3\right) + \mathscr{C}_{20} \left(I_1-3\right)^2 ,
	\label{WMRY}
\end{equation}
at the same order of approximation \cite{destrade10c}, with the partial derivatives $W_i$ of Eq.~\eqref{Elast} given by
\begin{equation}
	W^\text{MRY}_1 = \mathscr{C}_{10} + 2 \mathscr{C}_{20} \left(I_1-3\right) ,
	\qquad W^\text{MRY}_2 = \mathscr{C}_{01} .
	\label{WiMRY}
\end{equation}
The Rivlin coefficients $\mathscr{C}_{ij}$ are linked to the material parameters of fourth-order elasticity \eqref{WZabo} according to \cite{destrade10b}
\begin{equation}
	\begin{aligned}
		\mathscr{C}_{10} &= \tfrac12\left(2\mu + \tfrac14 \mathscr{A}\right) , &
		\mathscr{C}_{01} &= -\tfrac12\left(\mu + \tfrac14 \mathscr{A}\right) , &
		\mathscr{C}_{20} &= \tfrac14\left(\mu + \tfrac12\mathscr{A} + \mathscr{D} \right) , \\
		\mu &= 2 \left(\mathscr{C}_{10}+\mathscr{C}_{01}\right) , &
		\mathscr{A} &= -8 \left(\mathscr{C}_{10}+2\mathscr{C}_{01}\right) , &
		\mathscr{D} &= 2 \left(\mathscr{C}_{10}+3\mathscr{C}_{01}+2\mathscr{C}_{20}\right) ,
	\end{aligned}
	\label{ElasticConst}
\end{equation}
where the coefficients $\mathscr{C}_{10}$, $\mathscr{C}_{01}$ are the Mooney parameters, and $\mathscr{C}_{20}$ is the Yeoh parameter. If the only non-zero coefficient in \eqref{WMRY} is $\mathscr{C}_{10} = \mu/2$, then we recover the neo-Hookean model. Note that a more general second-order Rivlin series expansion can be linked to \eqref{WZabo} as well \cite{destrade10c}. The choice of suitable parameter values is described in the next subsection.

Let us summarise the equations of motion.
The governing equations are the balance laws \eqref{ConsF}-\eqref{ConsMom} to which we add the constitutive law \eqref{Constitutive}, the evolution of the memory variables \eqref{MemEvol} and the incompressibility constraint \eqref{Dilat}-\eqref{IncompDiff}. In three spatial dimensions and with a set of $N$ memory variables, we therefore end up with a system of $6N+13$ equations of the variables $\lbrace \bm{F}, \bm{v}, \bm{S}_1^\text{v}, \dots , \bm{S}_N^\text{v}, q \rbrace$. The latter is closed by the provision of appropriate boundary conditions.

\subsection{Illustration: antiplane shearing motions}\label{ssec:Transverse}

Similarly to Tripathi et al. \cite{tripathi19}, let us consider antiplane shearing motion such that $\bm{u} = u(X,Y,t)\, \bm{e}_3$ and
\begin{equation}
	\bm{F} = \begin{bmatrix}
		1 & 0 & 0\\
		0  & 1 & 0\\
		\partial_X u & \partial_Y u & 1
	\end{bmatrix} .
	\label{Deformation2D}
\end{equation}
Thus, the deformation is purely isochoric, and the incompressibility constraint is always satisfied. The equations of motion become
\begin{equation}
	\left\lbrace
	\begin{aligned}
		&\partial_t F_{31} + \partial_X (-v) = 0, \quad \partial_t F_{32} + \partial_Y (-v) = 0, \\
		&\partial_X P_{11} + \partial_Y P_{12} = 0, \quad \partial_X P_{21} + \partial_Y P_{22} = 0,\\
		&\partial_t v + \partial_X\big({-P_{31}}/\rho_0\big) + \partial_Y\big({-P_{32}}/\rho_0\big) = 0, \\
		&\partial_t [S^\text{v}_\ell]_{3\bullet} = \omega_\ell \left( g_\ell [S^\text{e}_\text{D}]_{3\bullet} - [S^\text{v}_\ell]_{3\bullet} \right) .
	\end{aligned}
	\right.
	\label{Sys2D}
\end{equation}

Let us linearise the equations of motion about an undeformed equilibrium state by following the steps in Berjamin and De Pascalis \cite{berjamin22c}. Assuming neo-Hookean behaviour $\mathscr{C}_{01} = \mathscr{C}_{20} = 0$, the constitutive law is modified as follows in the infinitesimal strain limit:
\begin{equation}
	\bm{P} = -q\bm{I} + 2\mu\, \bm{\varepsilon} - \sum_{\ell=1}^N \bm{S}_\ell^\text{v} ,\qquad
	\bm{S}^\text{e}_\text{D} = 2 \mu\left( \bm{\varepsilon} - \tfrac13 \text{tr}(\bm{\varepsilon}) \bm{I} \right) ,
	\label{Linear}
\end{equation}
where $\bm{\varepsilon} = \frac12 \left(\nabla \bm{u} + \nabla^\top \bm{u}\right)$ is the infinitesimal strain tensor, that is, the symmetric part $\text{sym}(\nabla \bm{u})$ of the displacement gradient tensor. Thus, the stress components in Eq.~\eqref{Sys2D} become $P_{11} = P_{22}\simeq -q$, $P_{12}=P_{21}\simeq 0$, $P_{3\bullet} \simeq \mu F_{3\bullet} - \sum_{\ell=1}^N [S^\text{v}_\ell]_{3\bullet}$, and $[S^\text{e}_\text{D}]_{3\bullet} \simeq \mu F_{3\bullet}$, where the bullet points represent indices in $\lbrace 1,2\rbrace$. Here, the hydrostatic pressure $q$ is a constant that equilibrates compressive tractions, i.e., $q\equiv 0$ if the material is stress-free at infinity. The remaining first-order system \eqref{Sys2D} is hyperbolic. Its characteristic wavespeeds along an arbitrary direction of propagation equal $\lbrace \pm c_\infty, 0 \rbrace$ where $c_\infty = \sqrt{\mu/\rho_0}$ is the speed of linear shear waves.

Harmonic wavefields $\propto \text{e}^{\text{i} (\omega t - \kappa X)}$ where $\text{i}$ is the imaginary unit must satisfy the \emph{dispersion relationship}
\begin{equation}
	\rho_0\frac{\omega^2}{\kappa^2} = \mu \left(1 - \sum_{\ell=1}^N \frac{g_\ell \omega_\ell}{\omega_\ell + \text{i} \omega}\right) ,
	\label{Dispersion2D}
\end{equation}
where $\omega = 2\pi f > 0$ is the angular frequency (in rad/s) and $\kappa$ in $\mathbb C$ is the wave number (in m\textsuperscript{$-1$}). In the high-frequency range, this dispersion relationship reduces to $\rho_0 \omega^2/\kappa^2 = \mu$, so that waves propagate at constant phase velocity $c_\infty$. In the low-frequency range, the phase velocity is constant as well, and it equals $c_0 = c_\infty \sqrt{1-\sum_\ell g_\ell}$.

\begin{figure}
	\centering
	\includegraphics{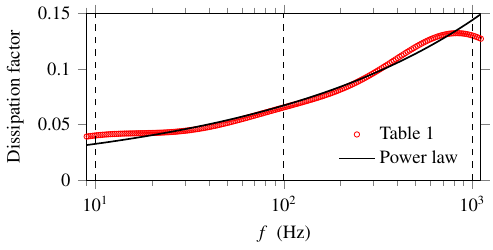}
	
	\caption{Dissipation factor $-\mathfrak{Im}(\kappa^2)/\mathfrak{Re}(\kappa^2)$ of infinitesimal harmonic waves deduced from Eq.~\eqref{Dispersion2D}, compared to the case of power-law attenuation \cite{tripathi19b}. The vertical dashed lines mark the relaxation frequencies $\omega_\ell/(2\pi)$ of Table~\ref{tab:Parameters}. \label{fig:Dispersion}}
\end{figure}

Based on measurements over the frequency range $75$~Hz-$675$~Hz, Tripathi et al.~\cite{tripathi19, tripathi19b} provide a set of parameter values for a gelatin sample, see Table~\ref{tab:Parameters}. The corresponding dissipation factor{\,---\,}i.e., the reciprocal of the quality factor{\,---\,}is represented in Figure~\ref{fig:Dispersion}, where comparison with the attenuation law $-\mathfrak{Im}\,\kappa = 0.0034\, \omega^{1.3}$ is provided.

\begin{table}
	\caption{Physical parameters of a gelatin sample with $N=3$ relaxation mechanisms \cite{tripathi19}. The shear-wave phase velocity $\omega/\mathfrak{Re}\, \kappa$ deduced from Eq.~\eqref{Dispersion2D} equals $1.42$~m/s at the frequency $75$~Hz, and $c_\infty \approx 1.638$~m/s at infinite frequency. \label{tab:Parameters}}
	\vspace{2pt}
	
	\centering
	{\renewcommand{\arraystretch}{1.1}
		\begin{tabular}{lc|lc|lc}
			\toprule
			$\rho_0$ [kg/m\textsuperscript{3}] & $10^3$ & $g_1$ & $0.0434$ & $\omega_1$ [rad/s] & $2\pi \times 10^1$ \\
			$\mu$ [kPa] & $2.684$ & $g_2$ & $0.0466$ & $\omega_2$ [rad/s] & $2\pi \times 10^2$\\
			$\beta$ & $4.4$ & $g_3$ & $0.2213$ & $\omega_3$ [rad/s] & $2\pi \times 10^3$ \\
			\bottomrule
	\end{tabular}}
\end{table}

In the infinitesimal strain limit, the constitutive model \eqref{Constitutive} is equivalent to Tripathi et al. \cite{tripathi19}. However, these models are not strictly equivalent in the nonlinear range. To link both models, we consider simple shearing motions $F_{32} \equiv 0$ in the low- and high-frequency elastic limits. The configuration is one-dimensional, i.e. all the spatial derivatives $\partial_Y$ in Eq.~\eqref{Sys2D} vanish. First, we consider extremely slow motions for which $\bm{S}_\ell^\text{v} \simeq g_\ell \bm{S}^\text{e}_\text{D}$ at all times (i.e., $\omega_\ell \to +\infty$). In this case, the equations of motion \eqref{Sys2D} involve the stresses
\begin{equation}
	P_{11}\simeq -q + 2\left(\mathscr{C}_{10}+2\mathscr{C}_{01}\right)\left(1 + F_{31}^2 \sum_{\ell=1}^N \frac{g_\ell}{3} \right) + 4 \mathscr{C}_{20} F_{31}^2 , \qquad
	P_{31} \simeq \left(1 - \sum_{\ell=1}^N g_\ell\right) \mu \left(F_{31} + \tfrac23\beta F_{31}^3\right) ,
	\label{Elastic1D}
\end{equation}
and $P_{21}\simeq 0$,
where nonlinear terms up to cubic order have been kept. The hydrostatic pressure is no longer uniformly constant, and the \emph{parameter of nonlinearity} $\beta$ in the expression of $P_{31}$ reads
\begin{equation}
	\beta = \frac{3 \, \mathscr{C}_{20}}{\mathscr{C}_{10} + \mathscr{C}_{01}} = \frac32 \frac{\mu + \mathscr{A}/2 + \mathscr{D}}{\mu} \, ,
	\label{Beta}
\end{equation}
which is the same expression as in the lossless case of Tripathi et al. \cite{tripathi19}. Now, if we consider extremely fast motions for which $\bm{S}_\ell^\text{v} \simeq \bm{0}$ at all times (i.e., $\omega_\ell \to 0$), then we recover Eqs.~\eqref{Elastic1D}-\eqref{Beta} with $g_\ell = 0$ for all $\ell$. A typical value of $\beta$ for gelatin is given in Table~\ref{tab:Parameters}.

Among the three elastic constants in \eqref{ElasticConst}, we have only access to two values based on the parameters $\mu$ and $\beta$ of Table~\ref{tab:Parameters}. To complete the modelling, we assume that $|\mathscr{C}_{01}| \ll |\mathscr{C}_{10}|$, which amounts to neglecting the second Mooney coefficient \cite{destrade10b}{\,---\,}thus the coefficients \eqref{WiMRY} simplify the elastic response \eqref{Elast} greatly. By setting $\mathscr{C}_{01} \simeq 0$, we keep only Yeoh-type constitutive behaviour with parameters $\mathscr{C}_{10}$, $\mathscr{C}_{20}$, so that \eqref{WMRY} can be linked to the Demiray--Fung strain energy function \cite{rashid14,jiang15}
\begin{equation}
	W^\text{DF} = \frac{3\mu}{4 \beta} \left(\text{e}^{\frac{2}3 \beta (I_1-3)} - 1\right) \simeq \frac{\mu}{2} \left( I_1-3 + \frac{\beta}3 (I_1-3)^2\right)
	\label{WFung}
\end{equation}
up to higher-order terms, with the connexion $\mu = 2\mathscr{C}_{10}$. Under the assumption $\mathscr{C}_{01} = 0$, the values of the elastic constants $\mu$ and $\beta$ in Table~\ref{tab:Parameters} produce the coefficients $\mathscr{C}_{10} = 1.342$~kPa and $\mathscr{C}_{20} =  1.968$~kPa.

\begin{remark}
	Even though the assumption $\mathscr{C}_{01} = 0$ is coherent with experimental literature based on the Demiray--Fung model \eqref{WFung}, it appears that this assumption is found less coherent with experimental works based on the fourth-order elasticity theory \eqref{WZabo}, see Table~1 of Destrade et al. \cite{destrade19} for related parameter values.
\end{remark}

\subsection{Properties}\label{ssec:PropHyp}

Let us consider an arbitrary deformation in three space dimensions of the form $\bm{u}(X,Y,Z,t)$. We rewrite the equations of motion in conservation form \cite{lee12,lee13}
\begin{equation}
	\partial_t {\bf q} + \partial_{X_I} {\bf f}^I({\bf q}, q) = {\bf r} ({\bf q}) \, , \qquad
	{\bf q} = \left[{\setlength{\arraycolsep}{0pt} \begin{array}{c}
			(F_{ij})\\
			\hline
			(v_i) \\
			\hline
			([S_1^\text{v}]_{ij})_{i\leqslant j} \\%
			\vdots \\%
			([S_N^\text{v}]_{ij})_{i\leqslant j}
	\end{array}}\right] ,
	\label{SysBalance}
\end{equation}
where the vector $\bf q$ gathers the components of the primary variables, and the incompressibility condition \eqref{IncompDiff} is enforced:
\begin{equation}
	\partial_{X_I} \varphi^{I}({\bf q}) = 0, \qquad \varphi^{I} = J F^{-1}_{Ij} v_j
	\label{SysIncomp}
\end{equation}
Summation over repeated indices is performed, where the spatial coordinates $X_{1\leqslant I\leqslant 3}$ equal $X$, $Y$, $Z$. The flux function components and relaxation function components
\begin{equation}
	{\bf f}^I = \left[{\setlength{\arraycolsep}{0pt} \begin{array}{c}
			(-v_i \delta_{j I}) \\
			\hline
			(-P_{iI}/\rho_0) \\
			\hline
			(0)\\
			\vdots \\
			(0)	
	\end{array}}\right] ,
	\qquad
	{\bf r} = \left[{\setlength{\arraycolsep}{0pt} \begin{array}{c}
			(0)\\
			\hline (0) \\
			\hline
			\omega_1\, (g_1 [S_\text{D}^\text{e}]_{ij} - [S_1^\text{v}]_{ij}) \\%
			\vdots \\%
			\omega_N\, (g_N [S_\text{D}^\text{e}]_{ij} - [S_N^\text{v}]_{ij})
	\end{array}}\right]
	\label{SysFunc}
\end{equation}
are deduced from Eqs.~\eqref{ConsF}-\eqref{ConsMom} and from the constitutive law \eqref{Constitutive}-\eqref{WMRY}. In the case of $N=3$ relaxation mechanisms, the above vectors have $6N+12 = 30$ components.

For later use, let us assume that the right-hand side ${\bf r}$ vanishes (i.e., we are in a high frequency range for which $\omega_\ell \to 0$). A plane wave solution of the form ${\bf q} = {\bf q}(\bm{N}\cdot\bm{X}- c t)$ with the hydrostatic pressure $q = q(\bm{N}\cdot\bm{X}- c t)$ propagates along the material direction oriented by the unit vector $\bm N$ at a characteristic speed $c$. The constrained system of balance laws \eqref{SysBalance}-\eqref{SysFunc} yields
\begin{equation}
	\left( N_I{\bf A}^I - c\, {\bf I} \right) {\bf q}' + q' N_I{\bf a}^I = {\bf 0} , \qquad
	N_I {\bf b}^I\cdot {\bf q}' = 0 ,
	\label{Eigval}
\end{equation}
where $\bf I$ denotes the identity matrix, and
\begin{equation}
	{\bf A}^I = \frac{\partial{\bf f}^I}{\partial {\bf q}} , \quad {\bf a}^I = \frac{\partial{\bf f}^I}{\partial q} , \quad {\bf b}^I = \frac{\partial \varphi^I}{\partial {\bf q}} ,
	\label{Jacob}
\end{equation}
see detailed expressions in the Appendix~\ref{app:Mat}.

In particular, the first lines of \eqref{Eigval}\textsubscript{1} entail the relationship $c\bm{F}' = -\bm{v}'\otimes \bm{N}$ between the components of ${\bf q}'$, whereas the last group of lines implies that the tensors $\bm{S}_\ell^{\text{v} \prime{}}$ vanish. Remaining lines then impose $\rho_0 c^2 \bm{v}' = -c\bm{P}'\bm{N}$, i.e.
\begin{equation}
	\rho_0 c^2 \bm{v}' = \bm{Q}\bm{v}' + c q' \bm{m}	, \qquad \bm{Q} = \bm{N}^\top \frac{\partial \bm{P}^\top}{\partial \bm F} \bm{N},
	\label{Acoustic}
\end{equation}
where $\bm Q$ is the \emph{acoustic tensor}, and $\bm{m} = \bm{F}^{-\top}\bm{N}$. 
Scalar multiplication of \eqref{Acoustic} by $\bm m$ yields the expression of $c q'$. Substitution in \eqref{Acoustic} then leads to the following generalised eigenvalue problem
\begin{equation}
	\rho_0 c^2 \left(\bm{I} - \bm{n}\otimes \bm{n}\right) \bm{v}' = \left(\bm{I} - \bm{n}\otimes \bm{n}\right) \bm{Q}\bm{v}' ,
	\label{SpeedGen}
\end{equation}
where $\bm{n} = {\bm m}/\|\bm{m}\|$ is a unit vector.
Using the above notations, Eq.~\eqref{Eigval}\textsubscript{2} imposes the orthogonality condition
\begin{equation}
	J \bm{m}^\top \bm{v}' = 0 .
	\label{Ortho}
\end{equation}
If $\bm{v}'$ is orthogonal to the vectors $\bm{m} \propto \bm{n}$, then Eq.~\eqref{SpeedGen} simplifies further as an eigenvalue problem of the form \cite{scott85}
\begin{equation}
	\rho_0 c^2 \bm{v}' = \bm{Q}^\text{s} \bm{v}' , \qquad
	\bm{Q}^\text{s} = \left(\bm{I} - \bm{n}\otimes \bm{n}\right) \bm{Q} \left(\bm{I} - \bm{n}\otimes \bm{n}\right) .
	\label{SpeedOrtho}
\end{equation}
Detailed expressions of the acoustic tensor and of the sound speeds are given in the Appendix~\ref{app:Mat}.

Conversely, if the fluxes ${\bf f}^I$ vanish and $\omega_\ell$ is arbitrary, then we are left with the time-domain ordinary differential equation $\partial_t {\bf q} = {\bf r} ({\bf q})$. In the vicinity of a state $\bf q$ that satisfies \eqref{SysIncomp}, this differential system is characterised by the eigenvalues of the Jacobian matrix ${\bf R} = {\partial{\bf r}}/{\partial {\bf q}}$. A look at the block structure of ${\bf R}$ shows that the eigenvalues belong to $\left\lbrace -\omega_N, \dots, -\omega_1 ,  0\right\rbrace$ where $\omega_\ell$ are the viscoelastic relaxation frequencies. Therefore, the relaxation matrix ${\bf R}$ is negative about any admissible state $\bf q$, and the frequencies $\omega_\ell$ describe the speed of relaxation towards $\bf q$.

\section{Numerical resolution}\label{sec:FV}

\subsection{Artificial compressibility}\label{ssec:Chorin}

To solve the present system numerically, we introduce an approximate incompressibility condition based on nearly-incompressible behaviour. For this purpose, let us consider weakly compressible QLV solids described by the strain energy function \cite{simo87}
\begin{equation}
	W = U(J) + \bar W(\bar{\bm C}) , \qquad
	U(J) = \tfrac12 K \left(J-1\right)^2,
	\label{Wcomp}
\end{equation}
where the volume-preserving part $\bar{\bm C} = J^{-2/3}\bm{C}$ of the right Cauchy--Green strain is unimodular. The volumetric strain energy $U$ is a function of the volume dilation $J = \det\bm{F} \not \equiv 1$, where $K = \lambda + \tfrac23 \mu > 0$ is the bulk modulus \cite{lee13} and $\lambda$ is the first Lam{\'e} coefficient. The bulk modulus takes large values in soft solids, i.e., we have $K \gg \mu$. In other words, the compressibility ratio $\epsilon = \mu/K$ is a small dimensionless parameter, $\epsilon \ll 1$. The function $\bar W$ describes the strain energy of incompressible solids such as Mooney--Rivlin--Yeoh materials \eqref{WMRY}. Alternative expressions of $U$ can be found in literature \cite{rossi16,castanar20}, some of which have striking advantages compared to the present quadratic function. For instance, contrary to other options, the function $U$ proposed in \eqref{Wcomp} does not require an infinite amount of energy to compress the body to a single point, a feature that contradicts basic continuum physics.

Following Simo \cite{simo87}, the viscoelastic constitutive law \eqref{Constitutive} now takes the form
\begin{equation}
	\bm{S} = JU' \bm{C}^{-1} + J^{-2/3}\, \text{Dev} \left( \bar{\bm S}^\text{e} - \sum_{\ell=1}^N \bm{S}^\text{v}_\ell \right)
	\label{ConstitutiveComp}
\end{equation}
with
\begin{equation}
	\bar{\bm S}^\text{e} = 2 \left(\bar W_1 + \bar I_1 \bar W_2\right) \bm{I} - 2\bar W_2 \bar{\bm C} ,\qquad
	\dot{\bm S}_\ell^\text{v} = \omega_\ell \left( g_\ell \bar{\bm S}^\text{e}_\text{D} - {\bm S}_\ell^\text{v} \right) ,
	\label{EvolComp}
\end{equation}
and $\bar{\bm S}^\text{e}_\text{D} = \text{Dev} ( \bar{\bm S}^\text{e})$, see notations in \eqref{Elast}.
The quantities $\bar I_i$ denote the principal invariants of the volume-preserving deformation $\bar{\bm C}$, and $\bar W_i = \partial \bar W/\partial \bar I_i$ is introduced in a similar fashion to the coefficients $W_i$ of Eq.~\eqref{Elast}.

Based on a suitable definition of $q$, we note that the initial constitutive model \eqref{Constitutive}-\eqref{MemEvol} is eventually recovered in the limit of perfect incompressibility $\epsilon\to 0$ where $J\to 1$ and $\bm{C} \to \bar{\bm C}$. If the pressure $p = -U'$ is introduced, then differentiation in time yields $\dot p = -U'' \dot J$, from which we deduce
\begin{equation}
	\partial_t p + K\, \nabla\cdot \left(J\bm{F}^{-1}\bm{v}\right) = 0 . 
	\label{Chorin}
\end{equation}
according to \eqref{IncompDiff} and \eqref{Wcomp}.
We note in passing that perfect incompressibility \eqref{Dilat}-\eqref{IncompDiff} is recovered at equilibrium $\partial_t p \simeq 0$.

This observation prompts us to note the following. If we expand all the unknown variables $\bm{F}$, $\bm{v}$, \dots, $p$ as power series of $\epsilon$ and inject this Ansatz in the equations of motion, then the leading-order terms correspond to the perfectly incompressible problem for which \eqref{IncompDiff} is enforced. The incompressible solution can then be used to estimate higher-order corrections for nonzero values of the small parameter $\epsilon$. This way, the compressible problem can be viewed as a perturbation of the incompressible one (see also Caforio and Imperiale \cite{caforio18}).

The above equation \eqref{Chorin} is reminiscent of the \emph{artificial compressibility} (AC) method proposed by Chorin \cite{chorin97} for steady incompressible flows, where $K$ corresponds to the AC parameter. This correspondence follows from the present definition of the pressure $p = -U'$ in terms of $J$ based on the quadratic expression \eqref{Wcomp}\textsubscript{2}. In the case of unsteady incompressible flow, the artificial compressibility approach is often combined with dual time-stepping, i.e., a pseudo-time derivative $\partial_\tau$ is added to each equation of motion and $\partial_t p$ is replaced by $\partial_\tau p$. This way, the unsteady incompressible problem can be viewed as a steady state with respect to $\tau$, and numerical integration towards equilibrium in pseudo-time is implemented after each iteration in physical time \cite{drikakis05}.

Alternatively, the unsteady problem with artificial compressibility can be solved directly in physical time $t$. This approach results in a compromise on the choice of the AC parameter. Indeed, large values of $K$ increase the physical accuracy of the approximate model with respect to perfect incompressibility, but numerical errors also increase due to the presence of fast artificial waves that require the use of small time steps for stability reasons \cite{madsen06}. In spite of these properties, we select this latter approach for which a dedicated numerical method is designed and analysed in what follows.

\subsection{Properties}\label{ssec:PropChorin}

To analyse the properties of the AC method \eqref{ConstitutiveComp}, let us follow similar steps to Section~\ref{ssec:PropHyp}. Here the pressure $p$ is function of the volume dilation $J$. Hence, there is no need for a Lagrange multiplier as the motion is unconstrained. We define the flux functions ${\bf f}^I({\bf q})$ in Eqs.~\eqref{SysBalance}-\eqref{SysFunc} accordingly (Appendix~\ref{app:MatChorin}), so that the equations of motion take the form of a system of balance laws. 
The relaxation spectrum $\left\lbrace -\omega_N, \dots, -\omega_1 ,  0\right\rbrace$ corresponding to the right-hand side of the AC system is unchanged. It remains to evaluate the speed $c$ of plane waves propagating along the direction oriented by the unit vector $\bm N$ when the system's right-hand side is neglected.

For the AC system, Eq.~\eqref{Eigval} reduces to $( N_I{\bf A}^I - c\, {\bf I}) {\bf q}' = {\bf 0}$, with formally the same definition of ${\bf A}^I$ as in Eq.~\eqref{Jacob}. The previous equation still implies that the identities $c\bm{F}' = -\bm{v}'\otimes \bm{N}$ and $\rho_0 c^2 \bm{v}' = -c\bm{P}'\bm{N}$ are satisfied, and that the tensors $\bm{S}_\ell^{\text{v} \prime{}}$ vanish. Thus, we recover Eq.~\eqref{Acoustic} where $q'$ is virtually replaced by zero, which provides an eigenvalue problem for the determination of the characteristic sound velocities. Detailed expressions of the acoustic tensor are given in the Appendix~\ref{app:MatChorin}, and the sound speeds are estimated therein as well.

\subsection{Illustration: linear waves}\label{ssec:DispersionChorin}

To further investigate the consequences of the artificial compressibility method \eqref{ConstitutiveComp}, let us consider the infinitesimal strain limit \eqref{Linear} such that
\begin{equation}
	\bm{P} = \lambda\, \text{tr}(\bm{\varepsilon}) \bm{I} + 2\mu\, \bm{\varepsilon} - \sum_{\ell=1}^N \bm{S}_\ell^\text{v} ,\qquad
	\bar{\bm S}^\text{e}_\text{D} = 2 \mu\left( \bm{\varepsilon} - \tfrac13 \text{tr}(\bm{\varepsilon}) \bm{I} \right) ,
	\label{LinearChorin}
\end{equation}
and the remaining equations of motion are unchanged. Here, we have used the fact that the tensors $\bm{S}_\ell^\text{v}$ are trace-free, a property that can be deduced from the evaluation of the trace in \eqref{EvolComp}\textsubscript{2}. Note the similarity between Eqs. \eqref{LinearChorin} and \eqref{Linear}, in which the Lagrange multiplier $q$ can be viewed as the undetermined limit of $-\lambda \, \text{tr}(\bm{\varepsilon})$ as $\epsilon \to 0$, where $\lambda$ becomes very large but $\text{tr}(\bm{\varepsilon})$ vanishes \cite{depascalis14}.

Phase velocity and attenuation can be deduced from dispersion analysis. For harmonic wavefields $\propto \text{e}^{\text{i} (\omega t - \kappa \bm{N}\cdot \bm{X})}$ of angular frequency $\omega$ and wave number $\kappa$, the linear equations of motion yield the algebraic problem
\begin{equation}
	\rho_0 \frac{\omega^2}{\kappa^2} \bm{v} = \mu\left(1 - \sum_{\ell=1}^N \frac{g_\ell \omega_\ell}{\omega_\ell + \text{i} \omega}\right) \left(\bm{I} + \tfrac13\bm{N}\otimes\bm{N}\right) \bm{v} + K\left(\bm{N}\otimes\bm{N}\right) \bm{v}
	\label{DispersionChorin}
\end{equation}
whose solutions $\bm{v} = \bm{v}_\perp + \bm{v}_\parallel$ with $\bm{v}_\perp \cdot \bm{N} = 0$ must satisfy
\begin{equation}
	\rho_0 \frac{\omega^2}{\kappa^2} \bm{v}_\perp = \mu \left(1 - \sum_{\ell=1}^N \frac{g_\ell \omega_\ell}{\omega_\ell + \text{i} \omega}\right) \bm{v}_\perp ,\qquad
	\rho_0 \frac{\omega^2}{\kappa^2} \bm{v}_\parallel = \mu \left( \frac43 \left(1 - \sum_{\ell=1}^N \frac{g_\ell \omega_\ell}{\omega_\ell + \text{i} \omega}\right) + \frac{1}{\epsilon} \right) \bm{v}_\parallel .
	\label{DispersionChorinOrtho}
\end{equation}
Therefore, transverse waves $\bm{v} = \bm{v}_\perp$ propagate according to the usual dispersion relationship \eqref{Dispersion2D}, whereas the speed of artificial longitudinal waves $\bm{v} = \bm{v}_\parallel$ depends on $\epsilon$.

In the limit of linear elasticity $\omega_\ell \to 0$, we recover the linear shear wave speeds $\pm c_\infty$, and the speed $\pm\sqrt{(\lambda+2\mu)/\rho_0}$ of compression waves. Note that the latter might become very large as $\epsilon \to 0$. The dissipation factor of compression waves reads
\begin{equation}
	-\frac{\mathfrak{Im}(\kappa^2)}{\mathfrak{Re}(\kappa^2)} = \frac{ \displaystyle\sum_{\ell=1}^N \frac{g_\ell\omega_\ell \omega}{\omega_\ell^2 + \omega^2} }{ \displaystyle 1 + \frac{3}{4\epsilon} - \sum_{\ell=1}^N \frac{g_\ell\omega_\ell^2}{\omega_\ell^2 + \omega^2} } .
\end{equation}
Thus, these waves propagate at infinite velocity and without attenuation in the limit of perfect incompressibility.

\subsection{A split finite volume scheme}\label{ssec:Scheme}

We consider a finite computational domain for $\bm X$ which is discretised using a regular grid with mesh size $\Delta x$ in the $X$-direction, $\Delta y$ in the $Y$-direction, and $\Delta z$ in the $Z$-direction. Grid node coordinates with integer indices $i$, $j$, $k$ are denoted by $\bm{X}_{i,j,k} = (i\, \Delta x, j\, \Delta y, k\, \Delta z)$. The nodes are located at the center of the finite volumes (or cells), whose dimensions are $\Delta x \times \Delta y \times \Delta z$. A variable time step $\Delta t = t_{n+1}-t_n$ is introduced. Therefore, ${\bf q}(\bm{X}_{i,j,k}, t_n)$ denotes the value of $\bf q$ at the grid node $\bm{X}_{i,j,k}$ and time $t_n$. Its numerical approximation is denoted ${\bf q}_{i,j,k}^n$. In what follows, the bullet point in ${\bf q}_{i,\bullet}^n$ is used as a placeholder for the indices $j$, $k$.

Usually, the derivation of finite volume methods involves spatial averaging of the equations of motion over a finite volume, or cell. In the present study, we will focus on finite volume schemes of order two or less. For such finite volume methods, the difference between the cell averages of $\bf q$ and its point values is of the same order as the scheme's accuracy, therefore the distinction between cell averages and point values is unnecessary \cite{leveque02}. As a result, the methods presented here have many formal similarities with finite difference methods.

Let us introduce a finite-volume implementation of the incompressible model with artificial compressibility \eqref{ConstitutiveComp}-\eqref{EvolComp}. To circumvent the stability restriction imposed by the right-hand side of \eqref{SysBalance} that stems from the viscoelastic material response (see Section~\ref{ssec:PropChorin}), we split the equations of motion \eqref{SysBalance} into two sub-systems \cite{lombard11}
\begin{equation}
	\begin{aligned}
		\text{(a)} \quad & \left\lbrace
		\begin{aligned}
			&\partial_t \bm{F} + \nabla(-\bm{v}) = \bm{0} ,\\
			&\partial_t \bm{v} + \nabla\cdot\big({-\bm{P}}/\rho_0\big) = \bm{0} ,\\
			&\partial_t \bm{S}_\ell^\text{v} = \bm{0}, \quad 1\leqslant \ell \leqslant N,
		\end{aligned}
		\right.
	\\
		\text{(b)} \quad & \left\lbrace
		\begin{aligned}
			&\partial_t \bm{F} = \bm{0} , \quad \partial_t \bm{v} = \bm{0}, \\
			&\partial_t \bm{S}_\ell^\text{v} = \omega_\ell\, \big(g_\ell \bar{\bm S}^\text{e}_\text{D} - \bm{S}^\text{v}_\ell\big) , \quad 1\leqslant \ell \leqslant N ,
		\end{aligned}
		\right.
	\end{aligned}
	\label{HSplit}
\end{equation}
whose numerical integration is alternated following a splitting procedure \cite{leveque02}. Note in passing that \eqref{HSplit}\textsubscript{a} is a first-order system of conservation laws where the memory variables $\bm{S}_\ell^\text{v}$ remain constant. Conversely, the variables $\bm{F}$, $\bm v$ and $\bm{S}_\text{D}^\text{e}$ are constant over time in the sub-system \eqref{HSplit}\textsubscript{b}, which amounts to a set of linear ordinary differential equations.
The speed of plane waves in \eqref{HSplit}\textsubscript{a} can be analysed in a similar fashion to Section~\ref{ssec:PropChorin}, see Eqs.~\eqref{SysBalance}-\eqref{EvolComp}. Detailed expressions of the acoustic tensor are given in the Appendix~\ref{app:MatChorin}.

At the computational level, each differential operator \eqref{HSplit} corresponds to a discrete operator $\mathcal{H}_a^{\Delta t}$, $\mathcal{H}_b^{\Delta t}$ for its numerical integration over one time step $\Delta t$. Here, we consider the second-order accurate \emph{Strang splitting} scheme \cite{leveque02}
\begin{equation}
	{\bf q}_{i,\bullet}^{n+1} = \mathcal{H}_b^{\Delta t/2} \mathcal{H}_a^{\Delta t} \mathcal{H}_b^{\Delta t/2} \,
	{\bf q}_{i,\bullet}^{n}
	\label{Splitting}
\end{equation}
corresponding to the integration of the full system based on the sub-systems \eqref{HSplit}.

In this study, we focus on robust numerical methods whose implementation and analysis are rather straightforward{\,---\,}a similar strategy was followed by others \cite{berjamin21b,favrie23}. Hence, the following choices are made:
\begin{itemize}
	\item For \eqref{HSplit}\textsubscript{a}, integration is performed numerically using a Godunov-type finite volume method based on dimensional splitting \cite{leveque02}, i.e. $\mathcal{H}_a^{\Delta t} = \mathcal{H}_\text{Z}^{\Delta t}\mathcal{H}_\text{Y}^{\Delta t}\mathcal{H}_\text{X}^{\Delta t}$ where the discrete operator $\mathcal{H}_\text{X}^{\Delta t}$ represents a spatially one-dimensional finite volume scheme along $X$. The corresponding time stepping formula is deduced from the Rusanov method (aka. local Lax--Friedrichs method, abbreviated LLF \cite{leveque02})
	\begin{equation}
		{\bf q}_{i,\bullet}^{n+1} =
		{\bf q}_{i,\bullet}^{n} - \frac{\Delta t}{\Delta x} \left( {\bm \Phi}_{i+\frac12,\bullet} - {\bm \Phi}_{i-\frac12,\bullet}\right) ,
		\qquad
		{\bm \Phi}_{i+\frac12,\bullet} = \frac12 \left( [{\bf f}^1]_{i,\bullet}^n +
		[{\bf f}^1]_{i+1,\bullet}^n  - \overline{c}_{i+\frac12,\bullet}  \left({\bf q}_{i+1,\bullet}^n - {\bf q}_{i,\bullet}^n\right) \right) ,
		\label{FVScheme}
	\end{equation}
	where the physical flux ${\bf f}^1$ is defined in Eq.~\eqref{SysFunc} with $I=1$. The scalar $\overline{c}_{i+\frac12,\bullet} = \max \big\lbrace |c|_{i,\bullet}, |c|_{i+1,\bullet} \big\rbrace$ approximates the maximum absolute wave speed at time $t_n$ about the cell interface $\bm{X}_{i+\frac12,\bullet}$, and the maximum value of $|c|$ is deduced from the eigenvalues of the acoustic tensor with $\bm{N} = \bm{e}_1$ (cf. Appendix~\ref{app:MatChorin}), which are evaluated numerically. The interface flux ${\bm \Phi}_{i-\frac12,\bullet}$ is calculated in a similar manner as above by performing the substitution $i \mapsto i-1$ in the formula that defines ${\bm \Phi}_{i+\frac12,\bullet}$. This step is stable under the Courant--Friedrichs--Lewy (CFL) condition
	\begin{equation}
		\Gamma = \overline{c} \frac{\Delta t}{\Delta x} \leqslant 1,
		\label{CFL}
	\end{equation} where $\Gamma$ is the Courant number along $X$, and $\overline{c} = \max_{i,\bullet} |c|_{i,\bullet}$ is the maximum absolute wave speed at time $t_n$ over the whole spatial domain. Similar time-stepping formulas $\mathcal{H}_\text{Y}^{\Delta t}$, $\mathcal{H}_\text{Z}^{\Delta t}$ are written for the directions $Y$, $Z$.
	\item For \eqref{HSplit}\textsubscript{b}, integration is performed in an exact fashion using the time stepping formula $\mathcal{H}_b^{\Delta t}$ defined by
	\begin{equation}
		[\bm{S}_\ell^\text{v}]_{i,\bullet}^{n+1} = \text{e}^{-\omega_\ell\Delta t} [\bm{S}_\ell^\text{v}]_{i,\bullet}^{n} + \left(1 - \text{e}^{-\omega_\ell\Delta t}\right) g_\ell [\bar{\bm S}_\text{D}^\text{e}]_{i,\bullet}^{n}
		\label{HRelNum}
	\end{equation}
	for all $\ell$. This step is unconditionally stable.\footnote{Note that the implementation proposed in Berjamin and Chockalingam \cite{berjamin21b} is based on an approximate formula.}
\end{itemize}
For the numerical method \eqref{FVScheme}, the calculation of ${\bf q}_{i,\bullet}^{n+1}$ requires the knowledge of ${\bf q}_{i-1,\bullet}^{n}$ and ${\bf q}_{i+1,\bullet}^{n}$. Therefore, to implement the present time-marching procedure, a dedicated procedure is required to update the data in the \emph{ghost cells}, which are located at the boundaries of the computational domain. Here, non-reflecting conditions are imposed at the boundaries, i.e., constant extrapolation in the ghost cells is performed at every time step (see Sections 7 and 21.8 of LeVeque \cite{leveque02}).

The method is implemented using the Julia programming language, and the resulting code is made available online at \href{https://github.com/harold-berjamin/SoftSol3D}{https://github.com/harold-berjamin/SoftSol3D}. The data ${\bf q}_{i,\bullet}^{n}$ at time $t_n$ is stored in three-dimensional arrays of vectors with $6N + 12$ components, where $N$ is the number of viscoelastic relaxation mechanisms. The procedure used to update the data from the time $t_n$ to $t_{n}+\Delta t$ is summarised below. Given ${\bf q}_{i,\bullet}^{n}$, the time step $\Delta t$, and known values of the parameters, we proceed as follows:
\begin{enumerate}
\item Update the data ${\bf q}_{i,\bullet}^{n} \leftarrow \mathcal{H}_b^{\Delta t/2} {\bf q}_{i,\bullet}^{n}$ according to \eqref{HRelNum}.
\item For each direction D = X, Y, Z,
\begin{enumerate}
	\item Calculate the interface fluxes ${\bm \Phi}$ along the relevant direction using \eqref{FVScheme}\textsubscript{2}.
	\item Update the data ${\bf q}_{i,\bullet}^{n} \leftarrow \mathcal{H}_{\text{D}}^{\Delta t} {\bf q}_{i,\bullet}^{n}$ according to \eqref{FVScheme}\textsubscript{1} by evaluating the flux difference along the relevant direction.
	\item Extrapolate the boundary data in the ghost cells along the relevant direction.
\end{enumerate}
\item Update the data ${\bf q}_{i,\bullet}^{n} \leftarrow \mathcal{H}_b^{\Delta t/2} {\bf q}_{i,\bullet}^{n}$ according to \eqref{HRelNum}.
\item Calculate the maximum absolute wave speed $\overline{c}$ by maximisation over the whole domain and over each direction based on the expression of the acoustic tensor in Appendix~\ref{app:MatChorin}.
\item Update the time step $\Delta t = \frac{\Gamma}{\overline{c}} \min\lbrace \Delta x, \Delta y, \Delta z\rbrace$, in agreement with \eqref{CFL}. 
\end{enumerate}
The steps 1.-3. correspond to the updating formula \eqref{Splitting}.

As we shall see, the speed $\overline{c}$ corresponding to fast compression waves can potentially be much larger than the shear wave speed, $c_\infty$, thus requiring that small time steps $\Delta t$ are chosen \eqref{CFL}. This feature and other properties will be analysed hereinafter in more detail. The method is modified further in Section~\ref{sec:Results} to reach nearly second-order accuracy in space and time.

\subsection{Properties}\label{ssec:Properties}

Let us consider a deformation gradient tensor of the form
\begin{equation}
	\bm{F} = \begin{bmatrix}
		1+\partial_X u_1 & 0 & 0\\
		\partial_X u_2 & 1 & 0\\
		\partial_X u_3 & 0 & 1
	\end{bmatrix} ,
	\label{F1D}
\end{equation}
which corresponds to a one-dimensional displacement field $\bm{u}(X,t)$. The exact motion is governed by the partial differential equations of Section~\ref{sec:GovEq}, and the constraint $\partial_X u_1 = 0$ is enforced ($\epsilon \to 0$). For the AC system, the finite volume approximation is defined in Eq.~\eqref{Splitting} with the sub-steps \eqref{FVScheme}-\eqref{HRelNum} and $\epsilon > 0$.

In this section, we consider the linear elastic limit for which $\omega_\ell \to 0$ and the displacement vector $\bm{u}(X,t)$ has infinitesimal components. Hence the analysis focuses on the linearised hyperbolic part \eqref{HSplit}\textsubscript{a}, which reduces to a set of linear conservation laws of the form $\partial_t {\bf Q} + {\bf M}\, \partial_X {\bf Q} = {\bf 0}$ with
\begin{equation}
	{\bf Q} = \begin{bmatrix}
		F_{I1}\\
		v_I
	\end{bmatrix} , \quad
	{\bf M} = \begin{bmatrix}
		0 & -1  \\
		-c_I^2 & 0 
	\end{bmatrix} , \quad
	\frac{c_I^2}{c_\infty^2} = 1 + \big(\tfrac13 + \tfrac1\epsilon\big) \delta_{I1} ,
	\label{SystDisp1D}
\end{equation}
for indices $I$ in $\lbrace 1, 2, 3\rbrace$. The case $I=1$ represents compression motions with longitudinal polarisation, whereas $I=2,3$ corresponds to shear motions with transverse polarisation. In the former case, the characteristic wave speed equals the speed $c_1 = \overline{c}$ of compression waves, whereas in the latter case, $c_{2,3} = c_\infty$ is the shear wave speed. Using Eq.~\eqref{SystDisp1D} with $I=1$, we note that $\overline{c}/c_\infty > \sqrt{4/3} \simeq 1.155$.

The scheme's time-stepping formula \eqref{FVScheme} reads
\begin{equation}
	{\bf Q}_i^{n+1} = \left(1-\Gamma\right) {\bf Q}_i^n - \tfrac{\Gamma}{2\overline{c}} {\bf M} \left({\bf Q}_{i+1}^n - {\bf Q}_{i-1}^n\right)  + \tfrac{\Gamma}{2} \left({\bf Q}_{i+1}^n + {\bf Q}_{i-1}^n\right)
	\label{FVDisp1D}
\end{equation}
where $\Gamma$ is the Courant number \eqref{CFL} and $\overline{c} = c_1$ is the maximum absolute eigenvalue of all the matrices ${\bf M}$. For a wave propagating at speed $c_I$, the effective Courant number equals $\Gamma \frac{c_I}{\overline{c}}$, which is necessarily smaller or equal to unity for all $I$. This observation implies that the enforcement of incompressibility in three dimensions leads to a diminution of the time step by a factor $c_\infty/\overline{c} < 0.866$ compared to the unconstrained 1D and 2D cases, to comply with the stability restriction \eqref{CFL} required by the present procedure. We examine the impact of this modification in terms of accuracy hereinafter.

\paragraph*{Numerical dispersion}

Injecting harmonic plane wave solutions $\propto \text{e}^{\text i (\omega t - \kappa X)}$ in the scheme's updating formula \eqref{FVDisp1D} entails the eigenvalue problem
\begin{equation}
	\xi {\bf Q}_i^n = \left[ \left(1-\Gamma+\Gamma\cos(\kappa\Delta x)\right) {\bf I} + \text{i} \tfrac{\Gamma}{\overline{c}} \sin(\kappa\Delta x) {\bf M} \right] {\bf Q}_i^n
	\label{NumDisp}
\end{equation}
where $\xi = \text{e}^{\text i \omega \Delta t}$ is the \emph{amplification factor} which arises in the Von Neumann stability analysis (see Section~IV.1.3 of the book by Godlewski and Raviart \cite{godlewski96}). The quantity $2\pi/(\kappa \Delta x)$ represents the number of points per wavelength for real wavenumbers $\kappa > 0$. The above algebraic problem yields the eigenvalues $\xi$ from which one deduces the real and imaginary parts of the angular frequency $\omega = \omega'+ \text i \omega''$ such that $\arg \xi = \omega' \Delta t$ and $-\ln |\xi| = \omega'' \Delta t$. Thus, the numerical phase velocity and attenuation coefficient are defined as $\omega'/\kappa$ and $\omega''$, respectively.

All the waves satisfy the same $2 \times 2$ eigenvalue problem \eqref{NumDisp}, which yields the amplification factor
\begin{equation}
	\xi = 1 - \Gamma + \Gamma \cos(\kappa \Delta x) \pm \text{i}\, \Gamma \tfrac{c_I}{\overline{c}}  \sin(\kappa \Delta x) .
	\label{Ampli}
\end{equation}
The amplification factor is solely a function of the global Courant number $\Gamma$, the effective Courant number $\Gamma \tfrac{c_I}{\overline{c}}$, and the number of points per wavelength (through $\kappa \Delta x$). This expression will now be used to evaluate the numerical phase velocity and attenuation coefficient.

If the AC parameter $\epsilon$ is assumed constant, then $c_I/\overline{c} \leq 1$ is constant. The present artificial compressibility method solves the differential system with approximate incompressibility up to some numerical error. In particular, the numerical phase velocity and wave attenuation deduced from the amplification factor \eqref{Ampli} can be approximated as follows:
\begin{equation}
	\frac{\pm\omega'}{\kappa c_I} \simeq 1 - \tfrac{1}{6} \left(1 - 3 \Gamma + 2 (\Gamma \tfrac{c_I}{\overline{c}})^2\right) (\kappa \Delta x)^2 ,
	\qquad
	\frac{\omega''}{\kappa c_I} \simeq \tfrac12 \left(\tfrac{\overline{c}}{c_I} - \Gamma \tfrac{c_I}{\overline{c}}\right) (\kappa \Delta x) ,
	\label{DispNumShearConst}
\end{equation}
for small $\kappa\Delta x$ when the AC parameter is constant. Thus, for constant Courant numbers, unattenuated waves of phase velocity $\pm c_I$ are recovered as the mesh is refined. Nevertheless, it is worth pointing out that for fixed $\Gamma \leq 1$, the coefficient in the expression of the numerical wave attenuation factor \eqref{DispNumShearConst}\textsubscript{2} increases with decreasing values of the effective Courant number $\Gamma \tfrac{c_I}{\overline{c}}$. This observation means that the speed ${\overline{c}}$ of artificial compression waves cannot be chosen arbitrarily large without penalising the attenuation of shear waves for $I = 2,3$.

Let us emphasize that this approach based on a constant value of the AC parameter does not ensure convergence towards the perfectly incompressible solution. Furthermore, a suitable choice of the AC parameter is not necessarily straightforward \cite{madsen06}. To overcome this issue, one might assume that the AC parameter is dependent on the mesh size. Typically, we set $\epsilon \propto (\Delta x)^\alpha$ for some non-negative exponent $\alpha$. With this assumption, it follows that $c_\infty/\overline{c}$ decays at the same speed as $(\Delta x)^{\alpha/2}$ when the mesh is refined.

Ideally, one would like to improve the numerical dispersion properties \eqref{DispNumShearConst} of shear waves based on this assumption. Unfortunately, this turns out to be impossible. In fact, the numerical shear wave attenuation is now of order $(\kappa \Delta x)^{1-\alpha/2}$, which decreases slower towards zero than for constant AC parameters, as can be seen from \eqref{DispNumShearConst}\textsubscript{2}. Therefore, it is not possible to fully mitigate the numerical errors of the method while enforcing convergence towards the truly incompressible solution.

From the above remarks, we notice also that $0\leq \alpha< 2$ is required to enforce convergence towards the truly incompressible solution with decaying numerical attenuation in shear. If $\alpha \simeq 0$ is chosen small, then numerical dispersion will not be extremely penalising, but convergence towards the truly incompressible solution will be slow. Conversely, if $\alpha \simeq 2$ is chosen large, then numerical dispersion will be very penalising, but convergence towards the truly incompressible solution will be fast. We analyse these features further in the next paragraph.

\paragraph*{Modified equations}

In complement to the above considerations, let us inject Taylor series expansions of the grid node values about ${X}_{i}$, $t_n$ in the scheme's time-stepping formula \eqref{FVDisp1D}. This process yields the \emph{equivalent system} (or \emph{modified system})
\begin{equation}
	\partial_t {\bf Q} + {\bf M}\, \partial_X {\bf Q} = \tfrac{1}{2} c_I \Delta x  \left( \tfrac{\overline{c}}{c_I}{\bf I} - \Gamma \tfrac{c_I}{\overline{c}} ({\bf M}/c_I)^2\right) \partial_{XX} {\bf Q}
	\label{EquivSys}
\end{equation}
at leading order, see Section~8.6 of the book by LeVeque \cite{leveque02} for the methodology. We point out that numerical dispersion properties \eqref{DispNumShearConst} could be deduced directly from \eqref{EquivSys} as well.

For shearing motions $I=2,3$, the first term in the right-hand side of \eqref{EquivSys} is of order $(\Delta x)^{1-\alpha/2}$, whereas the second term is of order $(\Delta x)^{1+\alpha/2}$. Therefore, the leading order of accuracy of the present AC numerical method with respect to the initial system is $1-\frac\alpha{2}$ in shear. This way, we recover the bounds $0 \leq \alpha < 2$ for the AC parameter's exponent. In addition, it appears that the relative truncation error is the same for all the components of $\bf Q$ at leading order in $\Delta x$, given that $({\bf M}/c_I)^2 = {\bf I}$ in the present case \eqref{SystDisp1D}. Regardless, the order of convergence is expected to be identical for all the components of $\bf Q$ according to Eq.~\eqref{EquivSys}.

The accuracy of other numerical methods is estimated in the Appendix~\ref{app:LW}. Therein, we show that the Godunov or upwind method does not suffer from the above restrictions. In fact, contrary to Rusanov's approximate Riemann solver \eqref{FVDisp1D}, the order of accuracy of the exact Riemann solver does not depend on $\alpha$. However, the implementation of the Godunov method is more involved for general nonlinear systems in multidimensional space \cite{leveque02}. Despite their convergence order is independent of $\alpha$, the higher-order finite difference methods of Lax--Wendroff or ADER type introduced in the Appendix~\ref{app:LW} are not well-suited for the computation of shock waves \cite{leveque02}. Therefore, we will keep the finite volume method \eqref{FVScheme} with local Lax--Friedrichs flux despite its limitations.

\begin{figure}
	\centering
	
	\begin{minipage}{0.5\textwidth}
		\centering
		
		\includegraphics{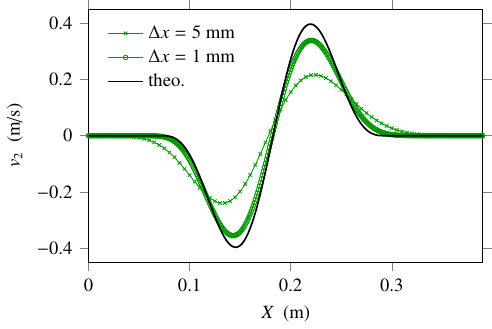}
		
		\vspace{0.2em}

	\end{minipage}\hfill
	\begin{minipage}{0.49\textwidth}
		\centering
		
		\includegraphics{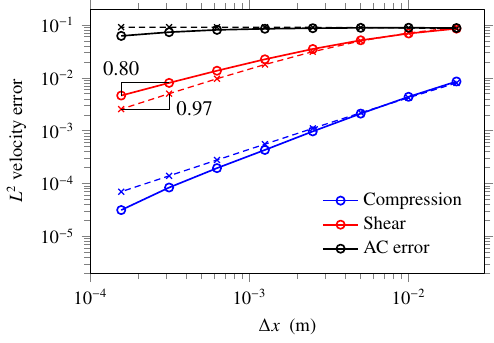}
		
	\end{minipage}
	
	\caption{Right-going shear waves in linear incompressible elasticity with sinusoidal forcing. Left: shearing velocity obtained numerically using the artificial compressibility method \eqref{FVDisp1D} for $\epsilon = 0.9$ and various mesh sizes $\Delta x$. Right: numerical error in $L^2$-norm with mesh-dependent AC parameter $\epsilon = 4\, (\Delta x/L)^{0.3}$ (solid lines) and with constant AC parameter $\epsilon = 0.9$ (dashed lines). \label{fig:Error1D}}
\end{figure}

Figure~\ref{fig:Error1D}-left displays numerical results obtained for several values of the mesh size $\Delta x$. Initially, the material is assumed undeformed and at rest. The spatial domain $-L/2 \leq X \leq L/2$ has length $L = 1$~m. Simulations were performed using \eqref{FVDisp1D} with $I \in \lbrace 1, 2\rbrace$ and $\Gamma = 0.95${\,---\,}we keep this value of the Courant number in later tests. A point source $s(t) \delta(X)$ with smooth sinusoidal signal
\begin{equation}
	s(t) = \sin(\Omega t) - \tfrac12 \sin(2\Omega t), \quad 0\leq \Omega t\leq 2\pi
	\label{SourceSig}
\end{equation}
in $\text{m}^2\text{s}^{-2}$ is added to the second line of the first-order system described by Eq.~\eqref{SystDisp1D} in the right-hand side. This way, we arrive at the non-homogenous wave equation
\begin{equation}
	\partial_{tt} v_I - c_I^2\, \partial_{XX} v_I = s'(t)\delta(X) , 
	\label{DAlembert}
\end{equation}
with no summation over repeated indices $I$. Consistently, the vector $s(t_{n+1}) \delta_{i0} \frac{\Delta t}{\Delta x}\, \bm{e}_2$ accounting for the source term is added to the expression \eqref{FVDisp1D} of ${\bf Q}_i^{n+1}$. Here, the fundamental angular frequency is set to $\Omega = 9\, \frac{\pi c_\infty}L \approx 46.3$~rad/s, and the shear wave speed $c_\infty$ is taken from Table~\ref{tab:Parameters}.

Figure~\ref{fig:Error1D}-right displays the evolution of numerical errors when the mesh is refined. The error is defined as the velocity error in $L^2$-norm between the numerical solution and a reference analytical solution at the final time $t \approx 0.18$~s for the same value of $\epsilon$. This error is evaluated numerically using Riemann sums over the domain $X\in \left]0.02, 0.47\right[$. Here, the analytical solution $v_I = \frac1{2c_I} s(t-|X|/c_I)$ to the partial differential equation \eqref{DAlembert} follows directly from d'Alembert's formula for inhomogeneous wave equations. As expected, the dashed curves produced for constant $\epsilon = 0.9$ show that the numerical method is first-order accurate. The solid curve obtained for mesh-dependent $\epsilon = 4\, (\Delta x/L)^{\alpha}$ with $\alpha=0.3$ illustrates the theoretical order of accuracy $1-\frac{\alpha}{2} = 0.85$ in shear. At the scale of the figure, this curve corresponds to values of $\epsilon$ ranging from 0.29 to 1.2. The slope of the compression error is not relevant given that these artificial waves exit the computational domain as $\epsilon$ becomes small, and since their amplitude $\propto 1 / \overline{c}$ decays towards zero.

In addition to these numerical errors, one might define the \emph{AC error} as the numerical error with respect to perfect incompressibility $\epsilon \to 0$ for $I=1$, which corresponds to the black lines of Figure~\ref{fig:Error1D}a-bottom. Again, the artificial compression waves exit the computational domain as $\epsilon \propto (\Delta x)^{0.3}$ becomes small. For constant $\epsilon = 0.9$, the AC error plateaus at small $\Delta x$ (dashed lines), thus highlighting some form of residual error which is characteristic of such an artificial compressibility method. To increase the accuracy globally, higher-order methods can be implemented, see for instance the next section and the Appendix~\ref{app:LW}. In addition, accuracy could be increased locally by implementing adaptive techniques.

\paragraph*{Volumetric locking}

In relation with incompressibility, \emph{volumetric locking} is a common feature of low-order finite element methods, for which the order of convergence is practically reduced to zero when the compressibility ratio $\epsilon$ becomes small \cite{suri96, lee17}. In this definition, the error to consider measures the accuracy of the numerical solution with respect to a reference solution obtained for the same constant value of $\epsilon$, or of Poisson's ratio, $\nu = \frac12 \frac{1- 2 \epsilon/3}{1+\epsilon/3}$. In this context, volumetric locking is caused by the inability of the approximate solution to comply with the constraints imposed by near incompressibility.

\begin{figure}
	\centering
	\includegraphics{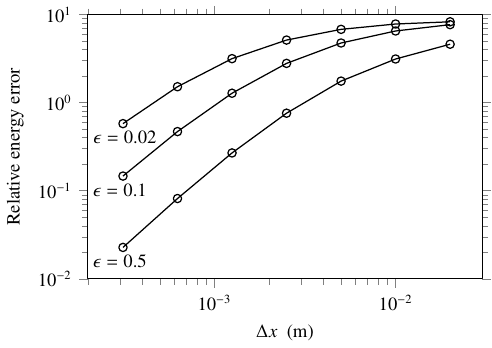}
	
	\caption{Evolution of the relative error in kinematic energy norm for several values of the compressibility ratio $\epsilon$. \label{fig:Locking}}
\end{figure}

Let us reconsider the problem \eqref{DAlembert} to investigate this issue. Thus, we estimate the relative error based on the kinematic energy norm $v_1^2+v_2^2$ evaluated using Riemann sums, in a similar fashion to the previous test. Several values of $\epsilon$ are considered, namely 0.02, 0.1, and 0.5, which correspond to Poisson ratios of 0.49, 0.45, and 0.36, respectively. The evolution of the relative error is displayed in Figure~\ref{fig:Locking}. The figure shows that while small values of the compressibility ratio $\epsilon$ have a detrimental impact on the resulting accuracy, the convergence speed remains nearly unaffected. Therefore, the present numerical method appears locking-free over the range of tested values of $\epsilon$, a result that is coherent with other works \cite{lee13}. Beyond the error curves shown in the figure, it is worth noting that extremely small values of $\epsilon$ might lead to plateauing error curves at the present mesh sizes.

\section{Numerical results}\label{sec:Results}

\subsection{MUSCL reconstruction}

To reach higher-order accuracy in the hyperbolic step \eqref{HSplit}\textsubscript{a}, the finite volume scheme \eqref{FVScheme} is improved by implementing MUSCL reconstruction in each direction with minmod-limited slopes \cite{toro09}. Nevertheless, we keep the same first-order dimensional splitting procedure for \eqref{FVScheme}, following the recommendations in Section~19.5 of LeVeque \cite{leveque02}. The procedure \eqref{HRelNum} for the integration of the relaxation step \eqref{HSplit}\textsubscript{b} is kept unchanged, as well as the second-order splitting scheme \eqref{Splitting}. Formally, the main algorithm described in Section~\ref{ssec:Scheme} is not modified, besides the computation of the numerical fluxes $\bm\Phi$ which now incorporates MUSCL reconstruction.

\begin{figure}
	\centering
	
	\begin{minipage}{0.5\textwidth}
		\centering
		
		Rusanov / LLF\vspace{0.1em}
		
		\includegraphics{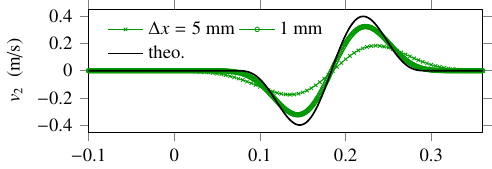}\vspace{-0.5em}
		
		MUSCL\vspace{0.1em}
		
		\includegraphics{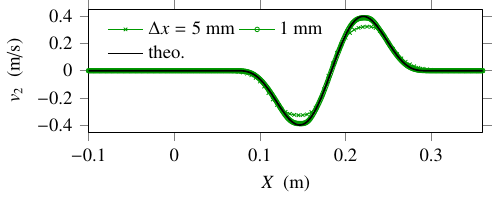}
	\end{minipage}\hfill
	\begin{minipage}{0.49\textwidth}
		\centering
		
		\includegraphics{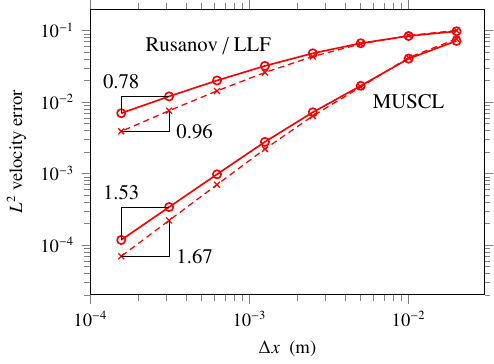}
	\end{minipage}
	
	\caption{Right-going shear waves in linear incompressible elasticity, Cauchy problem \eqref{CauchyProb}. Left: shearing velocities obtained numerically using the artificial compressibility method \eqref{FVDisp1D} for $\epsilon = 0.9$ and various mesh sizes $\Delta x$. Right: numerical error in $L^2$-norm with mesh-dependent AC parameter $\epsilon = 4\, (\Delta x/L)^{0.3}$ (solid lines) and with constant AC parameter $\epsilon = 0.9$ (dashed lines). \label{fig:Cauchy1D}}
\end{figure}

The effect of the MUSCL--Hancock reconstruction procedure is illustrated in Figure~\ref{fig:Cauchy1D}. Here, we solve a smooth Cauchy problem defined by the initial data
\begin{equation}
	 {\bf Q}(X,0) = \frac{s(-X/c_I)}{2 c_I} \begin{bmatrix}
		-1/c_I\\
		1
	\end{bmatrix}, 
	\label{CauchyProb}
\end{equation}
which produces right-going shear waves with the same waveform as previously for $I=2$. In fact, this problem is better suited for the estimation of the order of accuracy than the previous one \eqref{DAlembert}, in which the source term is singular.
Qualitatively, we note that the accuracy in shear is greatly improved{\,---\,}underestimated wave magnitudes in Figure~\ref{fig:Cauchy1D}-left with the MUSCL scheme are caused by the minmod limiter.

Estimations of the order of accuracy in shear are provided in Fig.~\ref{fig:Cauchy1D}-right. In effect, for the LLF scheme with constant $\epsilon = 0.9$ (or respectively, with mesh-dependent $\epsilon = 4\, (\Delta x/L)^{0.3}$), numerical simulations show that the order of accuracy in shear is estimated equal to 0.96 (respectively 0.78) without MUSCL reconstruction, and 1.67 (respectively 1.53) when MUSCL reconstruction is implemented. In terms of computational costs, we observe that for $\epsilon=0.9$ and $\Delta x = 2$~mm (or respectively, $\Delta x = 1$~mm), the numerical simulation takes about 30~s (respectively, about 1.5~min) without MUSCL reconstruction, and about 1~min (respectively, about 4~min) when MUSCL reconstruction is implemented. Thus, the simulation time increases quadratically with decreasing values of the mesh size, with and without MUSCL reconstruction. These results were obtained using a Julia code developed in-house and run on a laptop with CPU processor (Intel\textsuperscript{\textregistered} Core\textsuperscript{\texttrademark} i7-8665U, 1.9 GHz, 2.11 GHz).

As discussed in Section~\ref{ssec:Properties}, the time step $\Delta t$ decreases with decreasing values of the parameter $\epsilon$, thus leading to an increase of the computational costs to reach the final time. Furthermore, due to the increase of numerical errors associated with decreasing values of $\epsilon$ (see Figs.~\ref{fig:Error1D}-\ref{fig:Cauchy1D}), a finer mesh size $\Delta x$ is also required to reach the final time with a given accuracy of the numerical solution. Therefore, decreasing values of $\epsilon$ are associated with increasing computational costs, leading to a compromise for the choice of $\epsilon$. In the upcoming examples, we select moderate values of $\epsilon$ that lead to reasonable computational costs, and we display the artificial compression waves that are generated through this process. It turns out that the amplitude of these artificial compression waves remains much smaller than that of the shear waves. Finally, we conclude that the present 3D method has similar properties to the 1D scheme introduced by Berjamin and Chockalingam \cite{berjamin21b}, but that performance and accuracy are slightly penalised due to the enforcement of incompressibility.

\subsection{Linear viscoelastic case}

In this section, we consider the linear viscoelastic problem with sinusoidal forcing (see Section~\ref{ssec:Properties} for the definition of the acoustic source), whose quasi-analytical solution is computed in Fourier domain using similar steps to those in Appendix~D of Favrie et al. \cite{favrie15}:
\begin{equation}
	v_{I}(X,t) = \frac{1}{2 \pi} \int_0^\infty \mathfrak{Re} \left( \frac{\hat{s}(\omega)}{c_I(\omega)} \text{e}^{\text{i}\omega\, \left\lbrace t - |X|/c_I(\omega)\right\rbrace } \right) \text{d}\omega .
	\label{FourierSol}
\end{equation}
Here, the complex-valued wave speed $c_I$ with positive real part satisfies the dispersion relationship \eqref{DispersionChorin}
\begin{equation}
	\frac{c_I(\omega)^2}{c_\infty^2} = \left(1 - \sum_{\ell=1}^N \frac{g_\ell \omega_\ell}{\omega_\ell + \text{i} \omega}\right) \left(1 + \tfrac13\delta_{I1}\right) + \tfrac1\epsilon \delta_{I1} ,
	\label{FourierDisp}
\end{equation}
and $\hat{s}(\omega) = \int s(t)\, \text{e}^{-\text{i}\omega t} \text{d}t$ defines the time-domain Fourier transform of the source.
This expression reduces naturally to the d'Alembert solution of \eqref{DAlembert} when $c_I$ is real and independent of the angular frequency $\omega$. In practice, $\hat s$ is evaluated numerically using a fast Fourier transform (FFT) algorithm based on a discretization of $s$ with 500 points per period and zero-padding for at least 10 other periods (rounding up the total number of points to the nearest higher power of two), and the integral \eqref{FourierSol} is evaluated as an inverse FFT.

\begin{figure}
	\centering
	\includegraphics{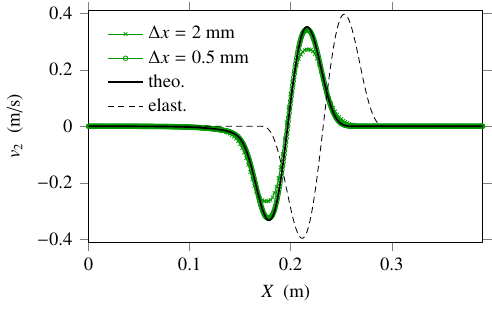}
	
	\caption{Right-going shear wave in linear incompressible viscoelasticity. Shearing velocity obtained numerically using the artificial compressibility method \eqref{FVDisp1D} with MUSCL reconstruction and $\epsilon = 0.9$ at various mesh sizes $\Delta x$. \label{fig:ErrorVisc}}
\end{figure}

Figure~\ref{fig:ErrorVisc} displays the linear viscoelastic solution with parameters from Table~\ref{tab:Parameters} obtained in a similar configuration as in Figure~\ref{fig:Error1D}b-top. The source has transverse polarisation along the $Y$-axis, with a loading angular frequency $\Omega$ of $16\, \frac{\pi c_\infty}L \approx 82.3$~rad/s and final time $t \approx 0.18$~s. At the computational level, the vector $s(t_{n+1}) \delta_{i0} \frac{\Delta t}{\Delta x} \bm{e}_{9+2}$ is added to the expression \eqref{Splitting} of the updated values ${\bf q}_{i,\bullet}^{n+1}$ after each iteration in time.

As illustrated in the figure by the obvious wave speed difference, the viscoelastic solid is less stiff than the elastic solid at the frequency of interest. This phenomenon is caused by wave dispersion properties, as shown through Eq.~\eqref{DispersionChorin}. Viscoelastic behaviour also causes the decay of wave amplitudes with increasing propagation distance (attenuation), among other dispersive effects. The figure shows the convergence of the present MUSCL scheme towards the analytical solution \eqref{FourierSol}.

\subsection{Nonlinear elastic case}

In the nonlinear elastic case, similar numerical simulations lead to amplitude-dependent wave distorsion caused by the generation of higher-order harmonics as the wave propagates. Furthermore, compression waves are produced through nonlinear coupling \cite{berjamin19,favrie23}. More precisely, if we expand all the unknown variables as power series of $\epsilon$ and inject this Ansatz in the equations of motion, we then find that the compression strain satisfies
\begin{equation}
	\partial_X u_1 \simeq \tfrac13 \epsilon \gamma^2 \left(1 + \tfrac23 \beta \gamma^2 \right)
	\label{Generation}
\end{equation}
at leading order in $\epsilon$, where $\gamma$ is the shear strain{\,---\,}for instance, the latter equals $\partial_X u_2$ if the point source $s(t) \delta(X)$ is polarised along the $Y$-axis, see the notations in Eq.~\eqref{F1D}.

\begin{figure}
	\centering
	
	\begin{minipage}{0.49\textwidth}
		\centering (a)
		
		\includegraphics{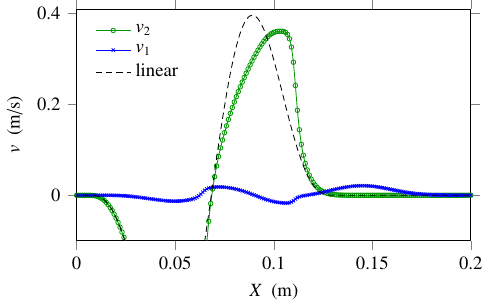}
	\end{minipage}\hfill
	\begin{minipage}{0.49\textwidth}
		\centering (b)
		
		\includegraphics{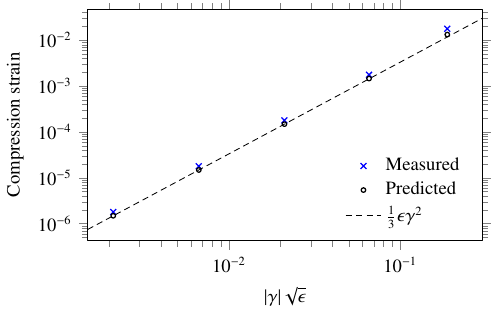}
	\end{minipage}
	
	\caption{Nonlinear incompressible elasticity. (a) Shear and compression velocities obtained numerically using the MUSCL scheme and $\epsilon = 0.8$ at the mesh size $\Delta x = 1$~mm. (b) Evolution of the coupling-induced compression wave amplitude in terms of a rescaled shear strain amplitude. \label{fig:Coupling}}
\end{figure}

We illustrate this property numerically in Figure~\ref{fig:Coupling}a where both the compression wave and the shear wave are represented. The configuration is similar to Figure~\ref{fig:ErrorVisc}, except that we now use a nonlinear elastic model with no relaxation mechanism ($N=0$), see Table~\ref{tab:Parameters} for the material parameters. Here, we have set $\epsilon = 0.8$, the mesh size satisfies $\Delta x = 1$~mm, and the source $s(t) \delta(X)$ is polarised along the $Y$-axis, see Eq.~\eqref{SourceSig} for the source signal. We have set $\Omega = 16\, \frac{\pi c_\infty}L \approx 82.3$~rad/s, and the waveforms are displayed at the final time $t\approx 0.08$~s.

Based on the above configuration, Figure~\ref{fig:Coupling}b represents the actual compression strain amplitude and the predicted amplitude \eqref{Generation} in terms of the rescaled shear strain magnitude $|\gamma|\sqrt{\epsilon}$. Amplitudes are estimated numerically based on the peak-to-peak magnitude at the final computational time. For each point in the figure, the source term $s(t) \delta(X)$ is multiplied by a given amplitude $A$ belonging to $[0.01,1]$, and estimated amplitudes are reported in the figure. We observe that the predicted trend is well reproduced. The slight mismatch between measured and predicted values is likely caused by higher-order terms $\propto \epsilon^2$ discarded in Eq.~\eqref{Generation}, as well as to numerical errors.

\subsection{Nonlinear viscoelastic case}

\begin{figure}
	\centering
	\includegraphics{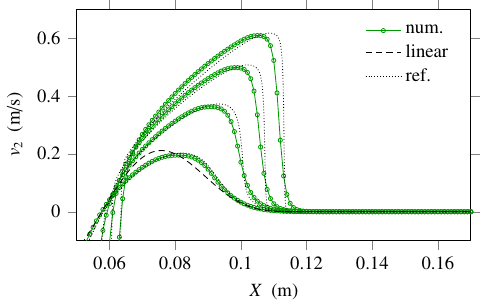}
	
	\caption{Nonlinear incompressible viscoelasticity. Shear velocity obtained numerically for several loading amplitudes using the MUSCL scheme with $\epsilon = 0.9$ and $\Delta x = 1$~mm. The reference is obtained with a one-dimensional code on a very fine grid \cite{berjamin21b} (dotted lines). \label{fig:EvolViscNL}}
\end{figure}

In the nonlinear viscoelastic case, wave propagation features illustrated above are combined, including wave dispersion, attenuation, nonlinear distorsion and compression-shear coupling. Here, we reconsider the configuration used in the previous test by setting all the material parameters to the values displayed in Table~\ref{tab:Parameters}. The source $s(t) \delta(X)$ polarised along $Y$ is multiplied by an amplitude $A$ belonging to $[0.5,2]$, and the shear velocities evaluated at the final time $t \approx 0.08$~s are reported in Figure~\ref{fig:EvolViscNL}. The numerical solution is obtained with $\epsilon = 0.9$ and $\Delta x = 1$~mm.

At large wave amplitudes, the formation of shear shock waves is clearly observed \cite{tripathi19}, although the discontinuities are more smeared out than in the highly-resolved 1D case \cite{berjamin21b} (dotted lines). These jumps in the shear velocity correspond to converging characteristics on each side of the singularity, i.e., the shear wave speed is larger on the wave crest that in an undeformed equilibrium state. Therefore, the observed strong gradients correspond to discontinuous shear wave solutions.

\subsection{Two-dimensional problem}\label{subsec:2D}

We consider a 0.6~m square of material centered at the origin of the coordinate system. An acoustic source with vertical polarisation is distributed along a vertical cylindrical surface of  circular cross-section, thus generating a telescopic shearing motion. The present configuration is very similar to that from the previous one-dimensional problem, except that the acoustic source distributed along the plane surface $X=0$ is now distributed along a curved surface of constant radius $R = 0.2$~m, where $R = \sqrt{X^2 + Y^2}$. Since this problem is invariant by translation along the vertical axis, we are facing a typical axisymmetric wave propagation problem.

Here, the loading signal \eqref{SourceSig} is multiplied by the amplitude $A = 0.005$. At the computational level, the vector
\begin{equation}
	s(t_{n+1}) \; \frac{\Delta t}{\Delta x \Delta y} \delta_{i\, i_0} \delta_{j \,j_0}
\end{equation}
is added to the expression \eqref{Splitting} of the updated values ${\bf q}_{i,\bullet}^{n+1}$ after each iteration in time. The sources are placed in the finite volume cells whose distance with the vertical cylinder of radius $R = 0.2$~m is minimal, and whose coordinates are the points ${\bm X}_{i_0, j_0, k}$. The domain is discretised using the uniform mesh size $\Delta x = \Delta y = 2$~mm. The AC parameter was set to $\epsilon = 0.9$, while the physical parameters are shown in Table~\ref{tab:Parameters}.

\begin{figure}
	\centering
	
	\begin{minipage}{0.49\textwidth}
		\centering (a) $t \simeq 0.06$~s
	\end{minipage}\hfill
	\begin{minipage}{0.49\textwidth}
		\centering (b) $t \simeq 0.12$~s
	\end{minipage}
	
	\includegraphics[width=0.9\textwidth]{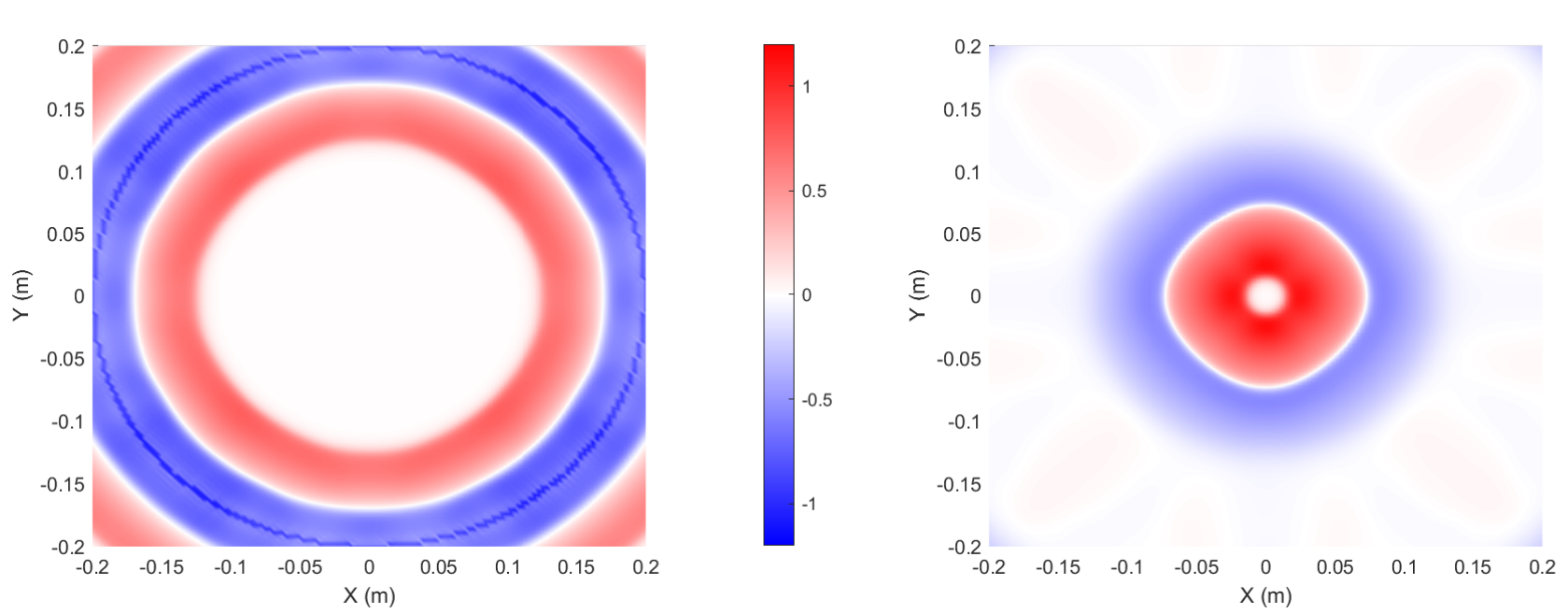}
	
	\vspace{0.5em}
	
	\begin{minipage}{0.49\textwidth}
		\centering		
		\includegraphics{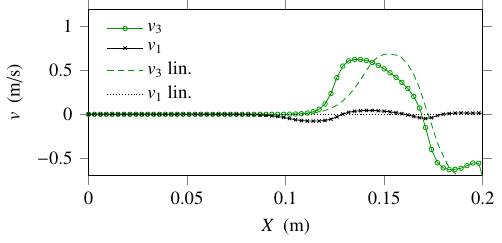}
	\end{minipage}\hfill
	\begin{minipage}{0.49\textwidth}
		\centering		
		\includegraphics{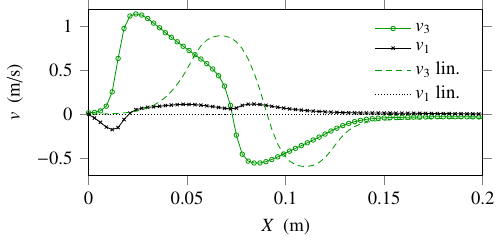}
	\end{minipage}
	
	\caption{Two-dimensional nonlinear viscoelastic problem. Top: snapshots of the velocity field $v_3$ with same polarisation as the source at increasing times. Bottom: evolution of the velocity components $v_3$, $v_1$ along the $X$-axis (solid lines with marks), and comparison with the linear solution ($\beta=0$, discontinuous lines). \label{fig:Visco2D} }
\end{figure}

Figure~\ref{fig:Visco2D} shows the results so-obtained. The heat maps of $v_3$ shown in the upper part of the figure were obtained by projection onto the plane $Z=0$. The evolution of the velocity components $v_3$ and $v_1$ along the $X$-axis is displayed underneath. As shown in the figure, the source generates artificial compression waves, visible in Figure~\ref{fig:Visco2D}-bottom where they correspond to the velocity field $v_1$ polarised along the $X$-axis. Besides the coupling between shear and compression \eqref{Generation} illustrated here, nonlinearity leads to the steepening of the wave fronts during propagation, and shear shock waves form. The increase of the wave amplitudes as they propagate towards the cylinder's axis is due to wave focusing.

\subsection{Three-dimensional problem}\label{subsec:3D}

To produce three-dimensional simulation results, we consider a 0.6~m cube of material centered at the origin of the coordinate system, which leads to the outward propagation of spherical waves. A truncated Gaussian source polarised along the $Y$-axis with sinusoidal loading is applied in the centre, where the angular frequency is $\Omega \approx 82.3$~rad/s and the final time is $t \approx 0.18$~s. The loading signal \eqref{SourceSig} is multiplied by the amplitude $A = 0.005$. At the computational level, the vector
\begin{equation}
	\Delta t\; s(t_{n+1}) \frac{\text{e}^{-{\|\bm{X}_{i,\bullet}\|^2}/{(2\sigma^2)}}}{(2\pi)^{3/2} \sigma^3} \frac{\mathbb{I}_{\lbrace \|\bm{X}_{i,\bullet}\|/\sigma < R \rbrace}}{\text{erf}(R/\sqrt{2}) - R \sqrt{2/\pi}\, \text{e}^{-R^2/2}} \bm{e}_{9+2} , \qquad R = 3.25,
\end{equation}
is added to the expression \eqref{Splitting} of the updated values ${\bf q}_{i,\bullet}^{n+1}$ after each iteration in time. The 3D spatial Gaussian function with standard deviation $\sigma$ is restricted to a sphere of radius $R \sigma$ centred at the origin by means of the indicator function, $\mathbb I$. The truncated distribution is normalised using a coefficient that involves the error function, $\text{erf}$.

\begin{figure}
	\centering
	
	\begin{minipage}{0.49\textwidth}
		\centering (a) $t \simeq 0.09$~s
	\end{minipage}\hfill
	\begin{minipage}{0.49\textwidth}
		\centering (b) $t \simeq 0.18$~s
	\end{minipage}
	
	\includegraphics[width=0.8\textwidth]{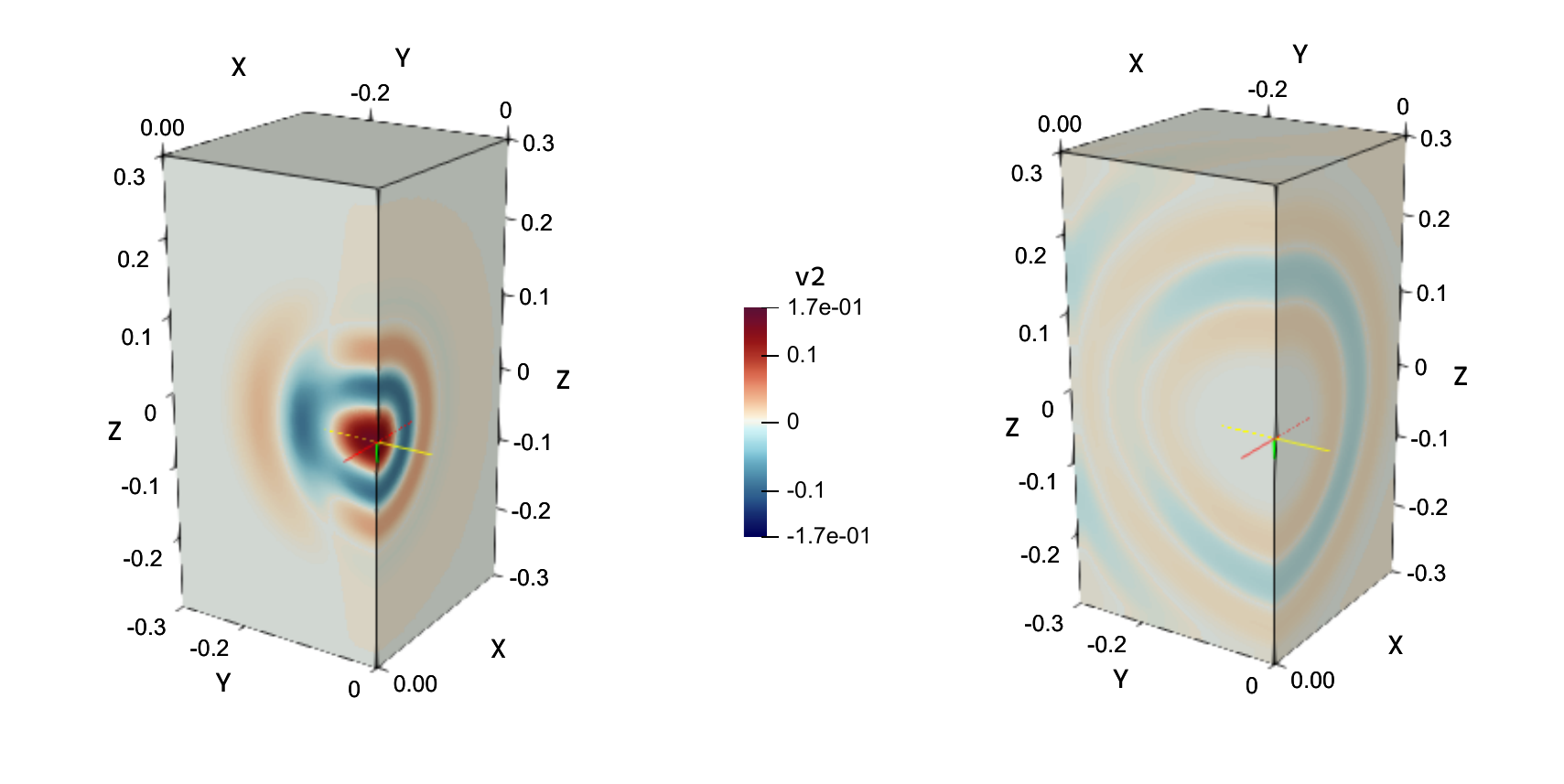}
	
	\begin{minipage}{0.49\textwidth}
		\centering		
		\includegraphics{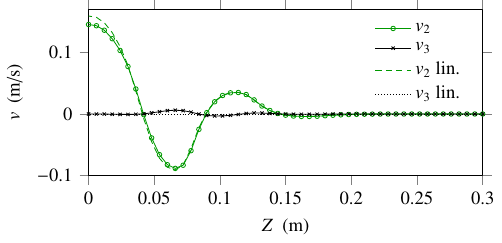}
	\end{minipage}\hfill
	\begin{minipage}{0.49\textwidth}
		\centering		
		\includegraphics{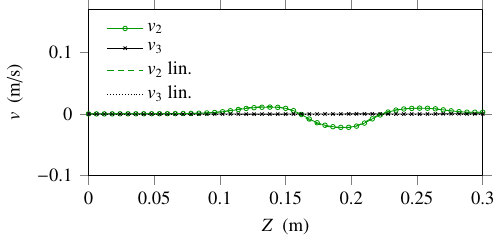}
	\end{minipage}
	
	\caption{Three-dimensional nonlinear viscoelastic problem. Top: snapshots of the velocity field $v_2$ with same polarisation as the source at increasing times. Bottom: evolution of the velocity components $v_2$, $v_3$ along the $Z$-axis (solid lines with marks), and comparison with the linear solution ($\beta=0$, discontinuous lines). \label{fig:Visco3D}}
\end{figure}

Figure~\ref{fig:Visco3D} shows the results so-obtained, where we have used the uniform mesh size $\Delta x = 6$~mm in every direction and the source width $\sigma = 18$~mm. The AC parameter was chosen in such a way that $\epsilon = 0.9$, while the physical parameters are set according to Table~\ref{tab:Parameters}. The heat maps of $v_2$ shown in the upper part of the figure were obtained by projection along the planes $X=0$ and $Y=0$. The evolution of the velocity components $v_2$, $v_3$ along the $Z$-axis is displayed underneath.

As shown in Fig.~\ref{fig:Visco3D}, the source generates artificial compression waves that propagate rapidly out of the domain. These waves are visible in Figure~\ref{fig:Visco3D}a-top, in the plane $X=0$. They can also be observed in the diagram below, where they correspond to the velocity field $v_3$ polarised along the $Z$-axis. Besides the coupling between shear and compression \eqref{Generation} illustrated here, nonlinearity does not have a very visible impact on wave propagation in the present case, and no shock wave forms. In fact, as displayed in Figure~\ref{fig:Visco3D}b, the shear waves and compression waves obtained in the nonlinear case ($\beta\neq 0$) and in the linear case ($\beta=4.4$) are nearly superposed. This is due to the effect of wave dispersion and of wavefront curvature, which both lead to a decrease of the amplitude of waves as they propagate.

Here we have illustrated the propagation of waves emitted by a spatially localised source, which is the contrary of wave focusing (Figure~\ref{fig:Visco2D}). Virtually, the effect of wave focusing can be inferred by reverting time in the above simulations. In this case, a particle velocity imposed at some distance from the centre concentrates at the origin with a large amplitude. This geometric effect could explain the damage of tissue observed in deep traumatic brain injuries.

\section{Conclusion}\label{sec:Conclusion}

A finite volume method based on MUSCL reconstruction is proposed for the numerical resolution of the equations governing nearly incompressible and truly incompressible solid dynamics. The material behaviour is described by Simo's theory of viscoelasticity \cite{simo87}, which is equivalent to Fung's quasi-linear viscoelasticity theory (QLV) in the incompressible limit. The numerical method is validated in the elastic case and in the infinitesimal-strain limit. Numerical results show that this method can be used for one-dimensional or multi-dimensional problems, but that numerical costs increase with the enforcement of incompressibility.

Computational costs could be mitigated by using MPI-based parallelisation techniques, or similar parallel computing environments. Moreover, based on the present approach, a higher accuracy could be obtained based on more sophisticated reconstruction techniques or other adaptive methods \cite{exahype}. Furthermore, the development of `pressure-Poisson' predictor-corrector schemes might be a promising venue for improved performance, see also Banks et al. \cite{banks23} for a finite difference procedure to solve the infinitesimal-strain case. Lastly, implementation on unstructured meshes will prove useful to address complex geometries such as actual brain models \cite{cinelli19}, see for instance the literature on Finite Element methods \cite{castanar23,lakiss23}.

\section*{Acknowledgment}

The author is grateful to Michael Dumbser (Trento, Italy) and colleagues for their hospitality and insightful advice, to Stephan Rudykh (Galway, Ireland) for support, as well as to Bharat B. Tripathi (Galway, Ireland) for helpful discussions. This project has received funding from the European Union's Horizon 2020 research and innovation programme under grant agreement TBI-WAVES{\,--\,}H2020-MSCA-IF-2020 project No 101023950.

\section*{Conflict of interest}

The author declares no potential conflict of interests.

\section*{Data availability statement}

The computer program used to produce the results can be accessed at \href{https://github.com/harold-berjamin/SoftSol3D}{https://github.com/harold-berjamin/SoftSol3D}.

\bibliography{biblio}{}

\appendix

\section{System coefficients}\label{app:Mat}

\subsection{Jacobian matrices}

Detailed expressions for Eq.~\eqref{Jacob} are given below. The Jacobian matrices ${\bf A}^I = {\partial{\bf f}^I}/{\partial {\bf q}}$ corresponding to the flux function ${\bf f}^I = -[v_i\delta_{jI},P_{iI}/\rho_0, \dots]^\top$ and the state vector ${\bf q} = [F_{kl},v_{k},\dots]^\top$ in Eqs.~\eqref{SysBalance}-\eqref{SysFunc} are given by
\begin{equation}
	-\rho_0 {\bf A}^I
	= \left[{\setlength{\arraycolsep}{1pt}
	\renewcommand{\arraystretch}{1.1}
	\begin{array}{c|c|ccc}
		(0) & (\rho_0\delta_{ik}\delta_{jI}) & (0) & \cdots & (0) \\
		\hline
		(\partial P_{iI}/\partial F_{kl}) & (0) & (-F_{ik}\delta_{Il}) & \cdots & (-F_{ik}\delta_{Il}) \\
		\hline
		(0) & (0) & (0) & \cdots & (0) \\
		\vdots & \vdots & \vdots & \vdots & \vdots \\
		(0) & (0) & (0) & \cdots & (0)
	\end{array}}\right] .
	\label{Jacobian}
\end{equation}
Related vectors defined by ${\bf a}^I = {\partial{\bf f}^I}/{\partial q}$ and ${\bf b}^I = {\partial \varphi^I}/{\partial {\bf q}}$ read
\begin{equation}
	{\bf a}^I = \left[{\setlength{\arraycolsep}{0pt} \begin{array}{c}
		(0)\\
		\hline
		(F^{-1}_{Ii}/\rho_0 )\\
		\hline
		(0) \\
		\vdots \\
		(0)
	\end{array}}\right] , \quad
	{\bf b}^I = \left[{\setlength{\arraycolsep}{0pt} \begin{array}{c}
			\big(J(F^{-1}_{lk}F^{-1}_{Ij} - F^{-1}_{Ik}F^{-1}_{lj}) v_j\big)\\
			\hline
			(JF^{-1}_{Ik}) \\
			\hline
			(0) \\
			\vdots \\
			(0)
	\end{array}}\right] .
	\label{Vectors}
\end{equation}

\subsection{Acoustic tensor}

For incompressible QLV materials, the first Piola--Kirchhoff stress tensor $\bm P$ is given by Eq.~\eqref{Constitutive}. In the case of Yeoh material, the elastic response \eqref{Elast} reads $\bm{S}^\text{e} = 2 W_1\bm{I}$ with the coefficient of Eq.~\eqref{WiMRY}\textsubscript{1}. 
The acoustic tensor $\bm Q$ defined in Eq.~\eqref{Acoustic} is given by
\begin{equation}
	\bm{Q} = q \bm{m} \otimes \bm{m} + a_0 \bm{I} + a_1 \bm{m}^\star \otimes \bm{m}^\star
	\label{AcousticTensor}
\end{equation}
with the coefficients
\begin{equation}
		a_0 = 2 W_1 - \bm{m}^\top\bm{\tau}^\text{v}\bm{m} , \quad 2W_1 = \mu \left(1 + \tfrac23\beta (I_1 - 3) \right) , \quad
		a_1 = 4 W'_1 = \tfrac43\mu\beta ,
	\label{AcousticTensorCoeffs}
\end{equation}
as well as the vectors $\bm{m} = \bm{F}^{-\top}\bm{N}$ and $\bm{m}^\star = \bm{F}\bm{N} = \bm{B m}$. The tensor $\bm{\tau}^\text{v} = \bm{F} (\sum_\ell \bm{S}_\ell^\text{v}) \bm{F}^\top$ represents the total viscous Kirchhoff stress, and the parameter of nonlinearity $\beta$ is defined in Eq.~\eqref{Beta}. In the infinitesimal strain limit, the acoustic tensor deduced from Eq.~\eqref{Linear} reads $\bm{Q} = \mu \left(\bm{I} + \bm{N} \otimes \bm{N}\right)$.

\subsection{Sound speeds}

The effective elastic modulus $\rho_0 c^2$ belongs to the spectrum of the tensor $\bm{Q}^\text{s}$ defined in Eq.~\eqref{SpeedOrtho}. Using the expression \eqref{AcousticTensor} of the acoustic tensor, we find
\begin{equation}
	\bm{Q}^\text{s} = a_0 \left(\bm{I} - \bm{n}\otimes\bm{n}\right) + a_1 \Lambda^{-2} \left(\bm{I} - \bm{n}\otimes\bm{n}\right) \bm{n}^\star \otimes \left(\bm{I} - \bm{n}\otimes\bm{n}\right) \bm{n}^\star ,
	\label{SpeedTensor}
\end{equation}
where we have introduced the coefficient $\|\bm{m}\|=1/\Lambda$ and the vectors $\bm{n}^\bullet = \bm{m}^\bullet/\|\bm{m}\|$. Thus, we arrive at
\begin{equation}
	\rho_0 c^2 \in \left\lbrace 0, a_0, a_0 + a_1 \Lambda^{-2} \sum_{j\neq I}B_{Ij}^2 \right\rbrace
	\label{SpeedSpeed}
\end{equation}
for waves propagating along the unit vector $\bm{n} = \bm{e}_I$. The null eigenvalue corresponds to compression waves polarised along $\bm n$, whereas the other eigenvalues correspond to shear waves polarised in the plane orthogonal to $\bm n$.

For instance, let us consider plane waves polarised along $\bm{e}_1$ and propagating along $\bm{n} = \bm{e}_2$ where the deformation gradient tensor $\bm{F} = \bm{I} + \gamma \bm{e}_1 \otimes \bm{e}_2$ describes simple shear. In this specific case, we have $\bm{N} = \bm{m} = \bm{e}_2$, and the relevant eigenvalue \eqref{SpeedSpeed} provides the expression
\begin{equation}
	\rho_0 c^2 = a_0 + a_1 \gamma^2 = \mu \left(1 + 2\beta \gamma^2 \right) - [\tau^\text{v}]_{22}
	\label{Chocka}
\end{equation}
for the shear wave speed $c$, which is also found in Eq.~(17) of Berjamin and Chockalingam \cite{berjamin21b}, up to the notations used therein. In the infinitesimal strain limit \eqref{Linear}, the tensor \eqref{SpeedTensor} reduces to $\mu \left(\bm{I} - \bm{n} \otimes \bm{n}\right)$, whose spectrum is the set $\lbrace \mu, 0 \rbrace$. As expected, we recover the nonzero sound velocities $\pm c_\infty$.

\section{AC system coefficients}\label{app:MatChorin}

\subsection{Jacobian matrices}

As specified in Section~\ref{ssec:PropChorin}, the flux ${\bf f}^I$ is now viewed as a function of ${\bf q}$ only. Formally, the expression \eqref{Jacobian} of the Jacobian matrix ${\bf A}^I$ is unchanged.

\subsection{Acoustic tensor}

The first Piola--Kirchhoff stress tensor $\bm{P} = \bm{F}\bm{S}$ is now given by Eq.~\eqref{ConstitutiveComp} with the isochoric elastic response \eqref{EvolComp} given by $\bar{\bm S}^\text{e} = 2 \bar W_1\bm{I}$. For convenience, we perform the substitution $\bar{\bm F} = J^{-1/3}\bm{F}$ in the expression of the stress. Upon differentiation with respect to $\bm F$ and projection along $\bm N$, we arrive at the following expression of the acoustic tensor \eqref{Acoustic}:
\begin{equation}
		\bm{Q} = q \bm{m} \otimes \bm{m}  + a_0 \bm{I} + a_1 \left(\bm{m}^\star - \tfrac13 I_1 \bm{m} \right) \otimes \left(\bm{m}^\star - \tfrac13 I_1 \bm{m} \right) 
		+ a_2\, \text{sym} \left(\bm{m} \otimes \left( 2\bar{W}_1 \bm{m}^\star - \bm{\tau}^\text{v} \bm{m} \right) \right) ,
	\label{AcousticTensorP}
\end{equation}
with the coefficients
\begin{equation}
	\begin{aligned}
		q\; &= KJ^2 + \tfrac59 J^{-\frac23} \left(2I_1\bar W_1 - \text{tr}({\bm\tau}^\text{v})\right) ,\quad
		a_0 = \left(2\bar W_1 - \bm{m}^\top{\bm \tau}^\text{v}\bm{m}\right) J^{-\frac23} , \quad 2\bar W_1 = \mu \left(1 + \tfrac23\beta (\bar I_1 - 3) \right) .\\
		a_1 &= 4 J^{-\frac43} \bar W'_1 = \tfrac43 J^{-\frac43} \mu \beta ,\quad
		a_2 = -\tfrac43 J^{-\frac23} .
	\end{aligned}
	\label{AcousticTensorCoeffsP}
\end{equation}
In the infinitesimal strain limit, the acoustic tensor deduced from Eq.~\eqref{LinearChorin} reads $\bm{Q} = \mu \bm{I} + (\lambda + \mu)\bm{N} \otimes \bm{N}$.

\subsection{Sound speeds}

The effective elastic modulus satisfies the eigenvalue problem $\rho_0 c^2 \bm{v}' = \bm{Q}\bm{v}'$ involving the acoustic tensor \eqref{AcousticTensorP}. In general, expressions of the sound speeds are too long to be displayed here. For simplicity, let us assume that $\bm N$ is a principal direction of the deformation (see Appendix~D of Lee \cite{lee12}), that is, we have $\bm{m}^\star = \Lambda^2 \bm{m}$ where $\Lambda = \|\bm{m}^\star\| = 1/\|\bm{m}\|$ denotes the principal stretch. Thus, we arrive at
\begin{equation}
	\bm{Q} = q \Lambda^{-2} \bm{n} \otimes \bm{n}  + a_0 \bm{I} + a_1 \left(1 - \tfrac13 I_1 \Lambda^{-2}\right)^2 \bm{n} \otimes \bm{n}
	 + 2\bar{W}_1 a_2 \bm{n} \otimes \bm{n} - a_2 \Lambda^{-2} \text{sym} \left(\bm{n} \otimes \bm{\tau}^\text{v} \bm{n} \right) .
	\label{AcousticTensPrinc}
\end{equation}
In particular, the elastic limit $\bm{\tau}^\text{v} = \bm{0}$ produces the sound speeds
\begin{equation}
	\rho_0 c^2 \in \left\lbrace a_0, a_0 + q \Lambda^{-2} + a_1 \left(1 - \tfrac13 I_1 \Lambda^{-2}\right)^2 + 2\bar{W}_1 a_2 \right\rbrace ,
\end{equation}
which recovers the neo-Hookean case studied in Lee et al. \cite{lee13} when $a_1=0$.
In the infinitesimal strain limit \eqref{LinearChorin}, the eigenvalues $\rho_0 c^2$ of the acoustic tensor belong to the set $\lbrace \mu, \lambda + 2 \mu \rbrace$, which is coherent with dispersion analysis \eqref{DispersionChorin}.

The expression of the acoustic tensor \eqref{AcousticTensPrinc} can be used directly for purely compressive motions. To do so, let us consider plane compression waves propagating along $\bm{n} = \bm{e}_1$ with $\bm{F} = \bm{I} + \varepsilon \bm{e}_1 \otimes \bm{e}_1$. In this case, we have $\bm{N} = J\bm{m} = \bm{e}_1$ where $J = 1+\varepsilon = \Lambda$, and the viscous stress tensor $\bm{\tau}^\text{v}$ is diagonal, see Eq.~\eqref{EvolComp}\textsubscript{2}. Thus, we find $\rho_0 c^2 = a_0$ or
\begin{equation}
	\rho_0 c^2 = a_0 + q \Lambda^{-2} + a_1 \left(1 - \tfrac13 I_1 \Lambda^{-2}\right)^2
	+ a_2 \left( 2\bar{W}_1 - \Lambda^{-2} [\tau^\text{v}]_{11} \right) ,
	\label{SpeedComp}
\end{equation}
which defines the compression wave speed along the $X$-axis.

\section{Other finite volume methods}\label{app:LW}

Let us consider generic finite-volume schemes of the form
\begin{equation}
	{\bf Q}_i^{n+1} = {\bf Q}_i^n - \tfrac{\Gamma}{2\overline{c}} {\bf M} \left({\bf Q}_{i+1}^n - {\bf Q}_{i-1}^n\right) + \tfrac{\Gamma c_I}{2 \overline{c}} {\bf D} \left({\bf Q}_{i+1}^n - 2{\bf Q}_{i}^n + {\bf Q}_{i-1}^n\right) ,
	\label{Disp1D}
\end{equation}
with a suitable expression of the discrete diffusion matrix ${\bf D}$. For instance, the Rusanov or local Lax--Friedrichs method \eqref{FVDisp1D} corresponds to the matrix ${\bf D} = \tfrac{\overline{c}}{c_I} {\bf I}$, see Table~\ref{tab:Accuracy}.
If we inject Taylor series expansions in the above time-stepping formula, then at leading order, we arrive at the modified system
\begin{equation}
	\partial_t {\bf Q} + {\bf M}\, \partial_X {\bf Q} = \tfrac{1}{2} c_I\Delta x \left( {\bf D} - \Gamma\tfrac{c_I}{\overline{c}} ({\bf M}/c_I)^2\right) \partial_{XX} {\bf Q} - \tfrac{1}{6} \Delta x^2 \left( {\bf I} - (\Gamma \tfrac{c_I}{\overline{c}})^2 ({\bf M}/c_I)^2\right) {\bf M}\, \partial_{XXX} {\bf Q} ,
	\label{DispModified1D}
\end{equation}
whose right-hand side yields the scheme's local truncation error. This expression encompasses the Rusanov method \eqref{EquivSys} as a special case. It can be simplified further by noting that $({\bf M}/c_I)^2 = {\bf I}$ in the present case \eqref{SystDisp1D}.

\begin{table}
	\caption{Modified equations \eqref{DispModified1D}-\eqref{DispModifiedADER} and order of accuracy for a class of numerical methods \eqref{Disp1D}. \label{tab:Accuracy}}
	\vspace{2pt}
	
	\centering
	{\renewcommand{\arraystretch}{1.3}
		\begin{tabular}{llc}
			\toprule
			Method & $\bf D$ & Order \\
			\midrule
			Lax--Friedrichs & $\frac{\overline{c}}{\Gamma c_I} {\bf I}$ & $1-\frac{\alpha}{2}$ \\
			Rusanov / LLF & $\frac{\overline{c}}{c_I} {\bf I}$ & $1-\frac{\alpha}{2}$ \\
			Godunov / Upwind & $|{\bf M}/c_I|$ & $1$ \\
			Lax--Wendroff & $\Gamma\tfrac{c_I}{\overline{c}} ({\bf M}/c_I)^2$ & $2$ \\
			Fourth-order ADER & n/a & 4 \\
			\bottomrule
	\end{tabular}}
\end{table}

Table~\ref{tab:Accuracy} summarises the diffusion and accuracy properties of various numerical methods, where the ratio $c_I/\overline{c}$ is assumed to decay at the same speed as $(\Delta x)^{\alpha/2}$ when the mesh is refined. Contrary to the Lax--Friedrichs method or to Rusanov's local Lax--Friedrichs approximate Riemann solver \eqref{FVDisp1D}, the order of accuracy of the exact upwind scheme does not depend on $\alpha$. The same observation applies to the second-order accurate Lax--Wendroff method.

For the sake of completeness, let us mention the case of a high-order generalisation of the Lax--Wendroff method known as the fourth-order ADER finite difference scheme \cite{lombard11}, which requires the vectors ${\bf Q}_{i-2}^n$, \dots, ${\bf Q}_{i+2}^n$ to calculate ${\bf Q}_i^{n+1}$. Compared to the previous numerical methods, the second-order accurate finite difference approximation of $\partial_X{\bf Q}$ in the right-hand side of \eqref{Disp1D} is replaced by a fourth-order accurate one, and further terms accounting for higher-order spatial derivatives are included. For this numerical method, the above analysis produces the modified system
\begin{equation}
	\partial_t {\bf Q} + {\bf M}\, \partial_X {\bf Q} = \tfrac1{6} \Delta x^4 \left(\tfrac1{5} {\bf I} - \tfrac1{4}(\Gamma \tfrac{c_I}{\overline{c}})^2 ({\bf M}/c_I)^2 + \tfrac1{20} (\Gamma \tfrac{c_I}{\overline{c}})^4 ({\bf M}/c_I)^4\right) {\bf M}\, \partial^{5}_{X}{\bf Q} ,
	\label{DispModifiedADER}
\end{equation}
at leading order of $\Delta x$, instead of \eqref{DispModified1D}. Therefore, the fourth order of accuracy of the ADER scheme is not impacted by the asymptotic enforcement of the incompressibility property, in a similar fashion to the second-order accuracy of the Lax--Wendroff method. Although these two numerical methods do not suffer from a reduction of the convergence order when the AC parameter varies with the mesh size, it should be emphasised that the magnitude of the local truncation error in shear is still affected by this process. Furthermore, contrary to the lower-order methods presented here, these higher-order finite difference methods will lead to the development of spurious oscillations near strong gradients \cite{leveque02}.

\end{document}